\documentclass[%
 reprint,
 amsmath,amssymb,
 aps,
]{revtex4-1}

\usepackage{latexsym}
\usepackage{textcomp}
\usepackage{graphics}
\usepackage{graphicx}

\begin{document}
\title{All-optical switching and strong coupling using tunable whispering-gallery-mode microresonators}

\author{D. O'Shea}
\affiliation{Institut f\"ur Physik, Johannes Gutenberg-Universit\"at Mainz, 55099 Mainz, Germany}
\affiliation{Vienna Center for Quantum Science and Technology, Atominstitut, TU-Wien, 1020 Vienna, Austria}
\author{C. Junge}
\affiliation{Institut f\"ur Physik, Johannes Gutenberg-Universit\"at Mainz, 55099 Mainz, Germany}
\affiliation{Vienna Center for Quantum Science and Technology, Atominstitut, TU-Wien, 1020 Vienna, Austria}
\author{M. P\"ollinger}
\affiliation{Institut f\"ur Physik, Johannes Gutenberg-Universit\"at Mainz, 55099 Mainz, Germany}
\author{A. Vogler}
\affiliation{Institut f\"ur Physik, Johannes Gutenberg-Universit\"at Mainz, 55099 Mainz, Germany}
\affiliation{Research Center OPTIMAS, Technische Universit\"at Kaiserslautern, 67663 Kaiserslautern, Germany}
\author{A. Rauschenbeutel}
\affiliation{Institut f\"ur Physik, Johannes Gutenberg-Universit\"at Mainz, 55099 Mainz, Germany}
\affiliation{Vienna Center for Quantum Science and Technology, Atominstitut, TU-Wien, 1020 Vienna, Austria}
\email{arno.rauschenbeutel@ati.ac.at}

\date{\today}
\begin{abstract}
We review our recent work on tunable, ultra-high quality factor whispering-gallery-mode bottle microresonators and highlight their applications in nonlinear optics and in quantum optics experiments. Our resonators combine ultra-high quality factors of up to $Q = 3.6\times10^8$, a small mode volume, and near-lossless fiber coupling, with a simple and customizable mode structure enabling full tunability. We study, theoretically and experimentally, nonlinear all-optical switching via the Kerr effect when the resonator is operated in an add--drop configuration. This allows us to optically route a single-wavelength cw optical signal between two fiber ports with high efficiency. Finally, we report on progress towards strong coupling of single rubidium atoms to an ultra-high $Q$ mode of an actively stabilized bottle microresonator.
\end{abstract}

\maketitle

\section{Introduction}
\label{intro}
Optical microresonators hold great potential for many fields of research and technology \cite{Vah03}. They  are used for filters and switches in optical communications \cite{Chu99AnE,Djordjev02,Alm04All}, bio-(chemical) sensing \cite{Armani07}, microlasers \cite{Ilc06}, as well as for cavity quantum electrodynamics (CQED) applications such as  single-photon sources \cite{Wal06} and interfaces for quantum communication \cite{Kim08}. All these applications rely on the spatial and temporal confinement of light by the resonator, characterized by the resonator's mode volume $V$ and its quality factor $Q$. For a given in-coupled power, the resulting intra-cavity intensity is then proportional to $Q/V$. This ratio defines a key figure relating the coupling strength between light and matter in the resonator to the dissipation rates of the coupled system. The highest values of $Q/V$ to date have been reached with whispering-gallery mode (WGM) microresonators \cite{Kip04Dem}. WGM microresonators are monolithic, micron-sized, dielectric structures in which the light is guided near the surface by continuous total internal reflection.  This extremely lossless mechanism of confinement enables the ultra-high quality factors (UHQs) to exceed $10^8$.

Standard fused silica WGM microresonators, like dielectric microspheres, microdisks, and microtoroidal resonators, typically confine light in a narrow ring along the equator of the structure near the resonator surface \cite{Mat06}. While such equatorial WGMs have the advantage of a small mode volume, they also exhibit a large frequency spacing between consecutive modes, which scales inversely with the resonator diameter. In conjunction with the limited tuning range of monolithic WGM microresonators, tuning of  ultra-high $Q/V$ equatorial modes to an arbitrary frequency has not been achieved to date. This impedes their use in a large class of applications that require a resonance of the microcavity to coincide with a predetermined frequency. For this reason, the WGM ``bottle microresonator'' has recently received considerable attention \cite{Kakarantzas01,Ward06,Strelow08,Warken08} because it provides a simple and customizable mode structure while maintaining a favorable $Q/V$-ratio \cite{Sum04Whi,Lou05Tun}. These devices combine ultra-high quality factors of up to $3.6\times 10^8$ or, equivalently, up to $2\times10^6$ in terms of finesse. Moreover, they exhibit a small mode volume and nearly lossless fiber coupling, characteristic of WGM resonators, while still enabling full tunability \cite{Poe09Ult}. This makes them ideal candidates for CQED experiments.

The ability to tune the resonance frequency of microresonators to other frequency-critical elements is especially important when developing sophisticated optical processing applications.  For example, many CQED quantum information protocols rely on mapping the quantum state of one two-level system onto an optical field and then transferring it to another remote two-level system. Such CQED quantum networks require multiple resonators to be mutually resonant and to be linked, e.g., by optical fibers \cite{Kim08}. In these situations, it is important to be able to not only tune the resonator but to also stabilize its resonance frequency to an external reference with a high degree of precision for extended periods of operation. The need for frequency stabilization in the case of ultra-high $Q$ WGM microresonators is made even more apparent by considering the typical situation where a temperature change of only 1~mK is enough to change the resonance frequency by one linewidth. Here, we describe a scheme used to stabilize an ultra-high $Q$ bottle microresonator to an atomic transition using the Pound--Drever--Hall technique while in a vacuum environment \cite{Osh10,Osh11}.

\begin{figure}
\centering
\includegraphics[width=0.45\textwidth]{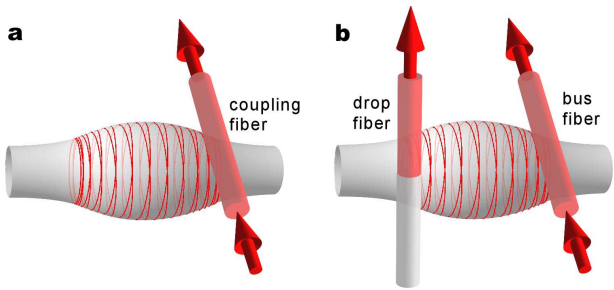}
    \caption{(a) Schematic of the coupling setup for exciting and spectroscopically characterizing resonator modes. A single ultra-thin coupling fiber is placed near the light field of the bottle microresonator mode. (b) Schematic of the coupling setup in the so-called ``add--drop'' configuration using two ultra-thin coupling fibers. The spiralling red line in the resonator traces the ray path.}
\label{bottle_scheme}
\end{figure}

Light can be coupled into and out of bottle microresonator modes with high efficiency by frustrated total internal reflection using the evanescent field of a sub-micron diameter tapered fiber coupler \cite{Kni97Pha,Spi03Ide}, see Fig.~\ref{bottle_scheme}~(a). Additionally, the bottle mode geometry offers simultaneous access to the resonator's light field with two coupling fibers, as shown in Fig.~\ref{bottle_scheme}~(b), without the spatial constraints inherent to equatorial WGMs. This facilitates the use of the bottle microresonator as a four-port device in a so-called ``add--drop configuration'' \cite{Poe10}. In communication technology, such devices are used for (de)-multiplexing optical signals. By judicious placement of the coupling fibers at the resonator caustics, we demonstrated the power transfer efficiency between the bus fiber and the drop fiber to be as high as 93~\% while, simultaneously, the filter bandwidth is as narrow as 49~MHz \cite{Poe10}. Importantly, these properties are compatible with strong coupling in a CQED experiment with neutral rubidium atoms.

Due to the strong enhancement of the intracavity intensity, microresonators are often used for non-linear optics applications. For example, they are employed for third harmonic generation \cite{Car07Vis} and the creation of frequency combs \cite{Hay07Opt}. Moreover, they are very interesting in the field of ``all-optical switching'', i.e., the control or redirection of a flow of light using a second light field. If the resonator material exhibits a third-order susceptibility, $\chi^{(3)}$, its refractive index $n$ depends on the intracavity intensity via the Kerr effect $n = n_{1} + n_{2} \times I$, where $n_1$ is the linear refractive index, $n_{\rm2} \propto \chi^{(3)} $ is the nonlinear refractive index, and $I$ is the intensity of the light field. A variation of the intra-cavity intensity then modifies the optical path length inside the resonator and thus changes its transmission properties. We recently demonstrated all optical switching via the Kerr effect at record-low powers using bottle microresonators \cite{Poe10}. Here, we present a simple model describing the optical switching behavior.

The article is organized as follows: In  Sec.~\ref{sec:Fabricate} we describe the fabrication of our ultra-thin optical fibers and bottle microresonators from commercial glass fibers using a heat-and-pull technique. In Sec.~\ref{sec:quality} we demonstrate ultra-high $Q$ factors with two complementary techniques: by recording the lifetime of the cavity photons and by performing spectroscopy of the resonator mode. The spectral properties of bottle modes and tunability are briefly described in Sec.~\ref{sec:Spectral}. An overview of the resonator optimization for nonlinear optics and quantum optics applications is presented in  Sec.~\ref{sec:Optmize}. The frequency stabilization of UHQ bottle microresonators in air and vacuum is presented and compared in  Sec.~\ref{sec:Stable}. In Sec.~\ref{sec:Switch} we theoretically study an experimental demonstration of optical bistability. An add--drop switch is realized by placing a second coupling fiber near the resonator light field. All-optical signal processing with bottle microresonators is demonstrated in Sec.~\ref{sec:All_optical}. An experiment to couple two spatially separated microresonators is presented in Sec.~\ref{sec:Separate}. Finally, in Sec.~\ref{CQED} we present an experiment that delivers cold rubidium atoms to the evanescent field of a bottle microresonator using an atomic fountain. The performance of the fountain is described and we give details of our bottle microresonator-based atom detection scheme.

\section{Fabrication of bottle microresonators and ultra-thin optical fibers}
\label{sec:Fabricate}
\begin{figure}
\centering
\includegraphics[width=0.45\textwidth]{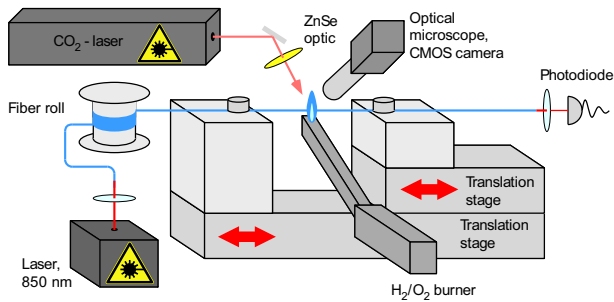}
    \caption{Schematic of the fiber pulling rig used to fabricate ultra-thin optical fibers and bottle microresonators. A commercial optical fiber is clamped to two translation stages. One stage is mounted on top of the other. The fiber is then heated by a hydrogen/oxygen flame with a width of 1~mm. The upper stage, called ``stretcher'', elongates the heated fiber, while the lower ``translator'' moves it relative to the flame (this method is commonly known as the ``heat-and-pull technique''). Alternatively, a focused CO$_2$-laser beam with a maximum power of 30~W (Synrad, Series 48-2) can be used as a heat source. The laser beam is focused by a ZnSe lens and only heats a 100~$\mu$m -- 150~$\mu$m wide section of the fiber. A microscope attached to a CMOS camera captures micrographs of the processed fibers. Throughout the ultra-thin optical fiber pulling process, the transmission of light from  a diode laser emitting at a wavelength of 850~nm is monitored.}
\label{pulling machine}
\end{figure}

Our ultra-thin optical fibers and bottle microresonators are produced from commercial glass fibers using a heat-and-pull technique. Typically, we use a step-index single mode fiber with an operation wavelength of 830~nm (Newport, F-SF). The company specifies a cladding diameter of 125~$\mu$m, a mode field diameter of 5.6~$\mu$m and an absorption of 5~dB/km. Before processing the fiber, its mechanical buffer is removed and the fiber surface is cleaned with acetone. The fiber pulling rig, used to produce both structures, is schematically shown in Fig.~\ref{pulling machine}.
A detailed description of the rig, as well as of the fabrication of ultra-thin optical fibers and bottle microresonators is given in \cite{Poe06,tesis_Warken}.

\subsection{Ultra-thin optical fibers}
Using a hydrogen/oxygen flame as a heat source, ultra-thin optical fibers with diameters down to 100~nm can be produced. In order to tailor the shape and slope of the taper transition and to create a homogenous waist of arbitrary length, the motion of the fiber relative to the heat source is controlled with translation stages. This motion is controlled via a computer and calculated to yield the desired fiber shape. In this work, we typically use tapered optical fibers (TOFs) with a waist diameter of 500~nm having a fabrication tolerance of around $\pm$3~\% of the target diameter. The transmission of a 852~nm laser beam through the TOF is monitored during the fabrication process as shown in Fig.~\ref{pulling machine} and the typical transmission after fabrication is greater than 97~\%.

\subsection{Bottle microresonators}
\label{fabrication_bottles}
\begin{figure}
 \centering
 \includegraphics[width=0.45\textwidth]{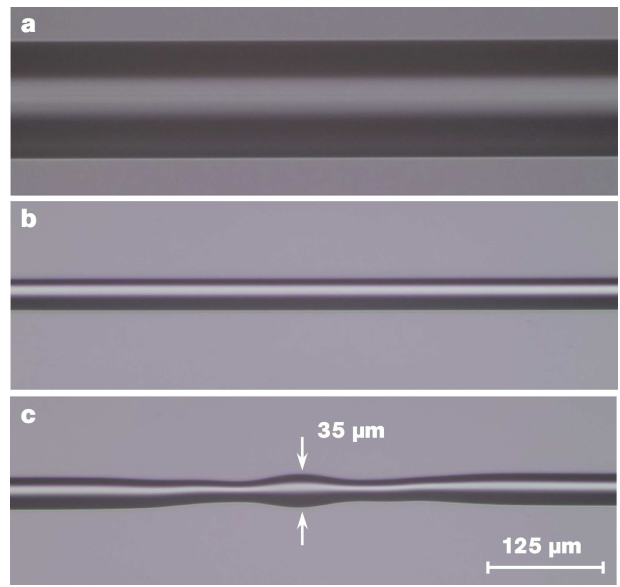}
    \caption{Micrographs showing the fabrication of a typical bottle microresonator in two steps, starting from (a) a commercial 125-$\mu$m diameter optical step-index fiber. (b) First, a section with a homogeneous diameter of 35~$\mu$m is created by simultaneously heating and stretching the fiber. (c) Next, local heating by a focused CO$_2$-laser beam and simultaneous elongation creates two microtapers on the fiber waist separated by approximately 150~$\mu$m. The bulge between the microtapers forms a bottle microresonator of diameter 35~$\mu$m.}
\label{steps}
\end{figure}

The fabrication process for bottle microresonators consists of two steps. Figure~\ref{steps} shows the structure resulting from each step. First, a uniform section of fiber with a length of a few millimeters and a diameter corresponding to the desired resonator diameter is fabricated via the heat-and-pull technique described above. This step is accomplished using the hydrogen/oxygen flame or a focused CO$_2$-laser beam. Next, a bulge between two microtapers is formed on the fiber waist. The microtapers are sequentially produced by locally heating the fiber waist with the focussed CO$_2$-laser beam, while slightly stretching the fiber. The central zone of the resulting bulge exhibits a parabolic variation of the fiber diameter and forms the bottle microresonator. Typical resonator diameters lie in the 30--45~$\mu$m range. Adjustment of the CO$_2$-laser beam spot size, the microtaper separation, and the elongation length allows one to precisely tailor the resonator geometry in order to obtain the desired spectral mode spacing.

\subsection{Techniques for estimating the curvature of bottle microresonators}
\label{sec:Curve}
The radius, $R$, of the highly prolate central region of the bottle around $z=0$ is well approximated by a parabolic profile
\begin{equation}
\label{radius profile} R(z)\approx R_0\cdot\left(1-\left(\Delta k\cdot z\right)^2/2\right).
\end{equation}
Here, $R_0$ is the maximum radius of the resonator at the position $z = 0$ and $\Delta k$ denotes the curvature of the resonator profile. Curvatures in the range of $\Delta k = 0.009$--0.020~$\mu$m$^{-1}$ are used in this work.

Following the resonator fabrication the curvature is measured with the aid of either an optical microscope image of the resonator itself ($\Delta k \gtrsim0.009$~$\mu$m$^{-1}$) or far-field diffraction of a laser beam ($\Delta k \lesssim0.009$~$\mu$m$^{-1}$).

\begin{figure}
 \centering
        \includegraphics[width=0.45\textwidth]{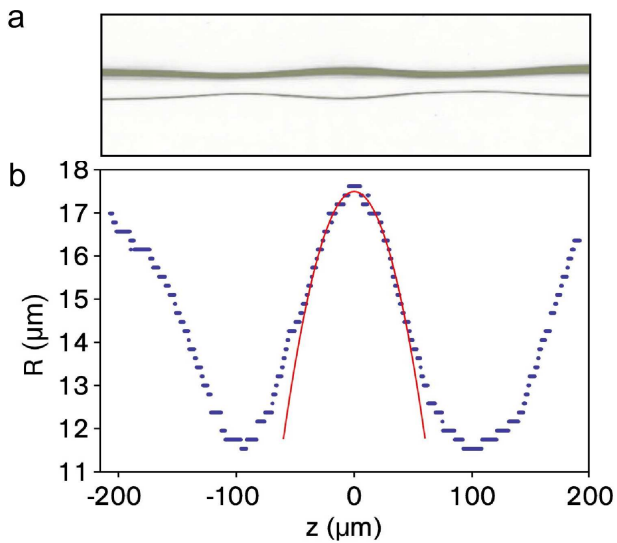}
\caption{Reconstructing the diameter profile of a bottle microresonator. (a) Micrograph of a typical bottle microresonator taken after fabrication. (b) Reconstructed radius profile with a parabolic fit (red) . Only the data in the region $z = \pm 35$~$\mu$m have been considered for the fit. In the central bulged region, where the mode is guided, the resonator profile shows close agreement with a parabola of the form $R(z)\approx R_0\cdot[1-\left(\Delta k\cdot z\right)^2/2]$. The fit yields $R_0 = 17.5\pm2.0$~$\mu$m and $\Delta k = 0.014\pm0.001$~$\mu$m$^{-1}$.}
\label{profil}
\end{figure}

A typical micrograph of a resonator and the inferred radius profile is shown in Fig.~\ref{profil}. The images are analyzed with a customized image analysis software. The program successively scans the pixels of each vertical line of the micrograph and automatically estimates the fiber edges. From this data, the profile can be calculated after calibration of the microscope using a test target with known dimensions. This method allows us to determine the local diameter with a precision of $\pm 2$~$\mu$m and $\Delta k$ with a precision of $\pm 0.001$~$\mu$m$^{-1}$. The accuracy is mostly limited by the resolution of the optical microscope and any possible saturation in the micrograph.

\begin{figure}
 \centering
        \includegraphics[width=0.45\textwidth]{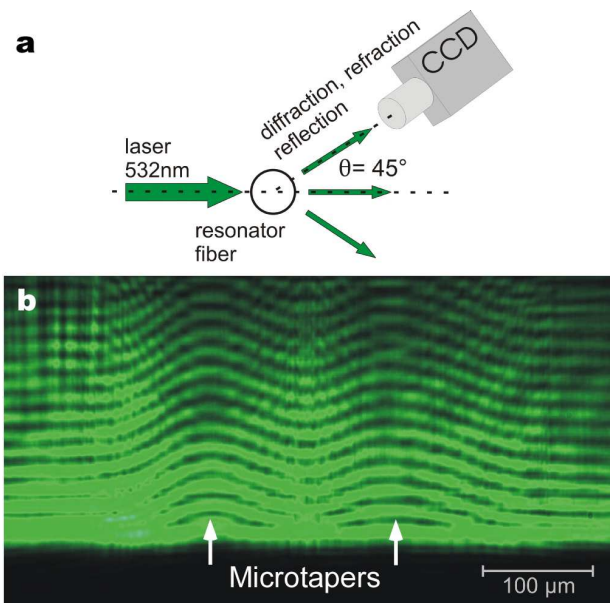}
\caption{(a) Setup of the scheme for determining the radius and curvature of bottle microresonators with low curvature ($\Delta k \lesssim0.009$~$\mu$m$^{-1}$). The resonator is illuminated with laser light at a wavelength of 532~nm and the resulting far-field diffraction is collected with a microscope at an angle of 45$^{\circ}$ relative to the incident beam. The microscope objective has a numerical aperture of 0.28 corresponding to a detection angle of 16$^{\circ}$. (b) The measured diffraction pattern reveals two regions where the diffraction pattern is modulated, indicating the microtapers on either side of the resonator. The difference in radius between the resonator region and the microtaper region is 330~nm while the curvature is $\Delta k = 0.0040 \mu$m$^{-1}$.}
\label{Beugung}
\end{figure}

The profile of bottle microresonators can also be reconstructed from a diffraction pattern (see Ref.~\cite{War04}). This is obtained by shining a collimated laser beam perpendicular to the resonator which, due to its small curvature, can be locally approximated as a cylinder. The dependency of the scattered intensity on the scattering angle and the fiber radius can then be easily calculated from Refs.~\cite{Poe06,Koz82,But98}. The resulting diffraction pattern is shown in Fig.~\ref{Beugung}~(b) and is obtained in the Fraunhofer regime. This method provides a radial resolution much better than 100~nm and is applicable to resonators having a curvature up to 0.009~$\mu$m$^{-1}$.

\section{Ultra-high Quality factors}
\label{sec:quality}
\begin{figure}[]
\centering\includegraphics[width=0.45\textwidth]{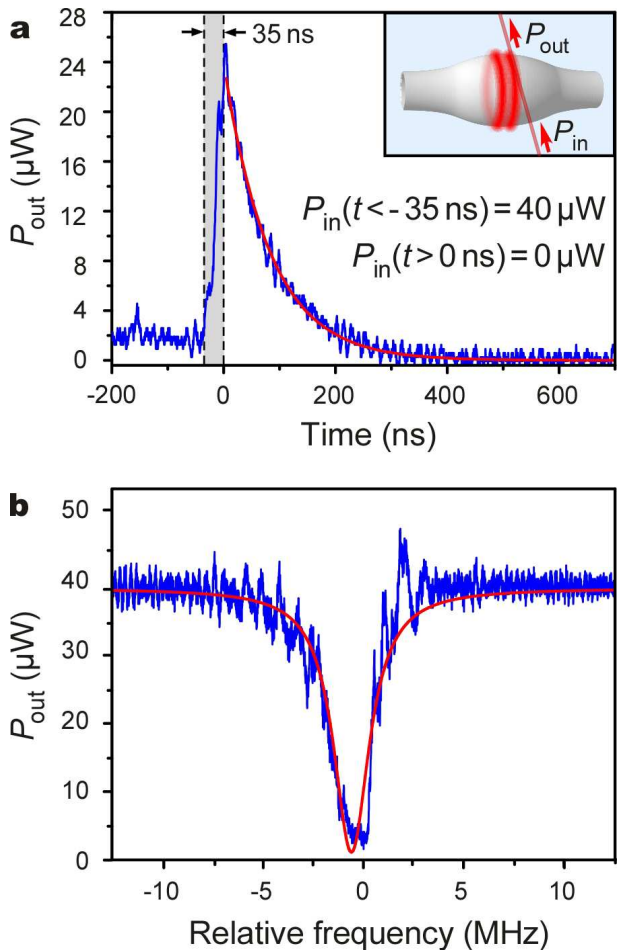}
    \caption{Photon lifetime in an ultra-high Q bottle microresonator. (a) Cavity ringdown measurement of a low order mode (axial quantum number $q = 1$, see Sec.~\ref{sec:Spectral}) in a bottle microresonator at critical coupling for a wavelength near 850~nm and transverse electric (TE) polarization, i.e., with the magnetic field parallel to the resonator surface. The photon lifetime, determined from the solid fitted curve, is  $\tau_{\rm crit} = 82$~ns, corresponding to an intrinsic quality factor in excess of $3.6 \times 10^8$. The inset schematically shows the coupling geometry with the sub-micron coupling fiber aligned with the light field of the bottle mode. (b) Spectrum of a separate bottle microresonator obtained at critical coupling. The solid line is a Lorentzian fit yielding a FWHM linewidth of $2.1\pm 0.1$~MHz, corresponding to an intrinsic quality factor in excess of $Q_0 = 3.3 \times 10^8$. The measured line is slightly asymmetric and broadened probably due to a combination of both thermal effects and interference between the cavity light and the quickly scanned laser frequency \cite{Mat06,Poi97}. This interference might also explain the ringing on the positive frequency side of the resonance.}
\label{Ringdown_linewidth}
\end{figure}

The quality factor of a resonator can be determined by measuring the photon lifetime $\tau$ and using the equation $Q = \omega\tau$, where $\omega = 2\pi\nu$ is the angular optical frequency. For this purpose, we use a cavity ringdown technique: by first critically coupling the resonator, we resonantly excite the bottle mode under investigation with the 850~nm probe laser. After switching off the probe beam within 35~ns using an acousto-optical modulator, the exponential decay of the intracavity power is monitored through the output port of the coupling fiber. This measurement, shown in Fig.~\ref{Ringdown_linewidth}~(a), is taken on a 35-$\mu$m diameter resonator with  $\Delta k = 0.012$~$\mu$m$^{-1}$ and yields a photon lifetime at critical coupling of $\tau_{\rm crit} = 82$~ns. Critical coupling is accomplished when the incident optical power is entirely dissipated in the resonator and the transmission of the coupling fiber at resonance drops to zero. We thus obtain a lower bound for the intrinsic photon lifetime in the uncoupled resonator of $\tau_0 = 2\tau_{\rm crit} = 164$~ns and an intrinsic quality factor in excess of $Q_0 = 3.6\times 10^8$. This ultra-high intrinsic quality factor is comparable to the values reported for other WGM microresonators of the same diameter \cite{Kip04Dem}. We note that at critical coupling, required for many applications including the CQED experiment described in Sec.~\ref{CQED}, our quality factor of $Q_{\rm crit} =  \omega\tau_{\rm crit} = 1.8\times10^8$ is about one order of magnitude larger than what has previously been reported in a suitable WGM microsonator \cite{Kip04Dem}. This ring-down measurement is independently confirmed by measuring the mode spectrum in a scheme wherein the frequency of the probe laser is scanned across the mode and the transmission is recorded. The spectrum of a second resonator with similar dimensions has a quality factor of $Q_0 = 3.3 \times 10^8$ and is shown in Fig.~\ref{Ringdown_linewidth}~(b). Several measurements on different resonators all show similar quality factors.

\section{Spectral properties and tunability}
\label{sec:Spectral}
\begin{figure}
\centering
\includegraphics[width=0.45\textwidth]{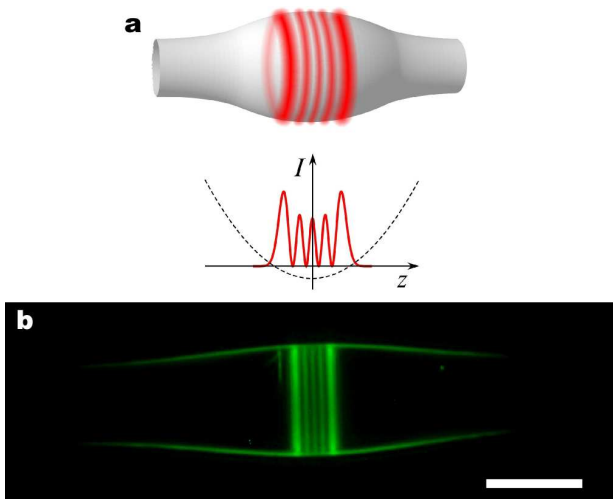}
    \caption{(a) Concept of the bottle microresonator. In addition to the radial confinement by continuous total internal reflection at the resonator surface, the axial confinement of the light is caused by a harmonic effective potential (dashed line) fixed by the curvature of the resonator profile. The resulting intensity distribution is therefore given by the eigenfunctions of the quantum mechanical harmonic oscillator \cite{Lou05Tun}. The intensity is significantly enhanced at the so-called ``caustics'' of the bottle mode, located at the classical turning points of the harmonic motion. This distinguishing feature gives bottle modes a three dimensional nature in comparison to equatorial WGMs. (b) Visualizing a $q=4$ bottle mode in a 36-$\mu$m diameter bottle microresonator. The bottle mode is excited by 850~nm laser light and visualized via the upconverted green fluorescence of dopant erbium ions. Scale bar, 30~$\mu$m.}
\label{bottle}
\end{figure}

Due to its highly prolate shape, the bottle microresonator gives rise to a class of WGMs with advantageous properties. The parabolic radius profile of the bottle microresonator in Fig.~\ref{bottle}~(a) defines a harmonic effective potential for the light field along the resonator axis. The light in these ``bottle modes'' harmonically oscillates back and forth along the resonator axis between two turning points which are defined by an angular momentum barrier \cite{Lou05Tun}.  The resulting axial standing wave structure exhibits a significantly enhanced intensity at the so-called ``caustics'' of the bottle mode, located at the turning points of the harmonic motion. These caustics therefore form two well separated coupling ports for coupling light or dipole emitters to the bottle mode. The mode structure is experimentally validated by visualizing the fluorescence from a resonator doped with erbium ions as shown in Fig.~\ref{bottle}~(b). Spectrally, the bottle microresonator possess equidistant eigenmodes, labeled by the ``azimuthal quantum number'' $m$, which counts the number of wavelengths that fit into the circumference of the  resonator, and the ``axial quantum number'' $q=0, 1, 2, 3,\ldots$, which counts the number of axial intensity nodes \cite{Sum04Whi,Lou05Tun}. The frequency spacing between modes with consecutive quantum numbers $q$ ($m$) is called the axial (azimuthal) free spectral range and will be denoted $\Delta\nu_q$ ($\Delta\nu_m$) in the following, cf. Fig.~\ref{tuning}. As in equatorial resonators, the azimuthal free spectral range is fixed by the resonator radius. However, the axial mode spacing, $\Delta\nu_q$, only depends on the curvature of the resonator profile, and can thus be made much smaller than $\Delta\nu_m$ without significantly affecting the mode volume of the resonator. It is therefore possible to couple light of any arbitrary frequency to the bottle microresonator by tuning the bottle resonance over one azimuthal free spectral range \cite{Lou05Tun,Poe09Ult}.

The azimuthal free spectral range defines the frequency spacing between neighboring mode families.  In small WGM resonators $\Delta\nu_m$ is typically very large. For example, changing $m$ by one (equal to fitting one extra wavelength into the circumference) for a 35-$\mu$m diameter WGM changes its resonance frequency by $\Delta\nu_m \approx 2$~THz, i.e., about one percent of the optical frequency. Tuning a WGM microresonator over such a large range could so far not be realized. The mode structure of bottle microresonators offers a simple route to achieving this goal and is described in the following \cite{Poe09Ult}.

\subsection{Mode spectrum}
The axial and azimuthal free spectral range $\quad \Delta\nu_q= \nu_{m,q+1}-\nu_{m,q}$ and $\Delta\nu_m=\nu_{m+1,q}-\nu_{m,q}$ can be derived from the eigenvalues $k_{m,q}$ of the wave equation $\label{} \left(\nabla^2 +  k^2\right)\cdot \vec{E} = 0$ \cite{Lou05Tun}. To a good approximation, they can be written as
\begin{equation}
\Delta\nu_m = \frac{c}{2\pi n} (k_{m+1,q}-k_{m,q})\approx \frac{c}{2\pi n R_0}
\label{azimuthal_FSR}
\end{equation}
and
\begin{equation}
\Delta\nu_q \approx \frac{c \Delta k}{2\pi n}~,
\label{axial_FSR}
\end{equation}
where $c/n$ is the speed of light in a medium with refractive index $n$. For a bottle microresonator with radius $R_0 = 17.5$~$\mu$m and curvature $\Delta k = 0.012$~$\mu$m$^{-1}$ the above formula yields an axial FSR (free spectral range) of $\Delta\nu_q = 391$~GHz that is about a factor of five smaller than the azimuthal FSR of $\Delta\nu_m = 1.9$~THz. Figure~\ref{tuning} schematically shows the spacing between the azimuthal FSR and the axial FSR for different axial modes $q$.

\subsection{Spectral tunability}
\begin{figure}
\centering
\includegraphics[width=0.45\textwidth]{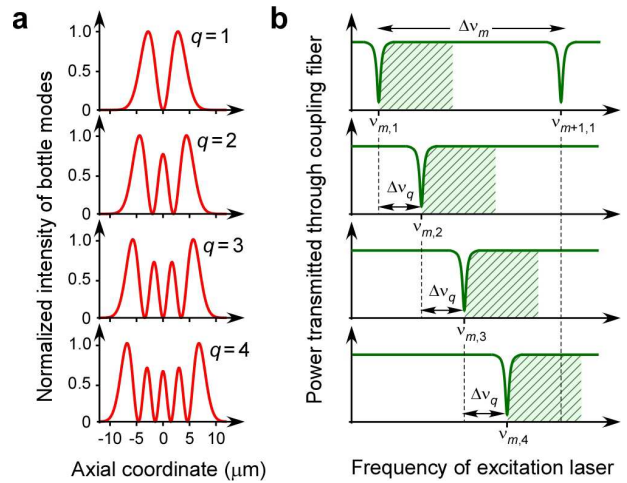}
    \caption{Tuning the bottle microresonator. (a) Calculated axial intensity distribution for bottle modes with axial quantum numbers $q = 1, 2, 3, 4$ in a bottle microresonator with 35-$\mu$m diameter and a curvature of  $\Delta k = 0.012~\mu$m$^{-1}$. (b)  Tuning scheme. Each of the bottle modes in (a) can be strain tuned by 700~GHz (shaded areas) which exceeds the free spectral range of $\Delta\nu_q = 395$~GHz between bottle modes of adjacent $q$ quantum numbers by a factor of 1.7. The resonance frequencies $\nu_{m,q}$ and $\nu_{m+1,q}$ of modes differing by one in the azimuthal quantum number $m$ are spaced by $\Delta\nu_m = 1.92$~THz. This frequency interval can thus be bridged by four consecutive axial modes, making any arbitrary frequency accessible by using the set of modes {$\nu_{m,1}$, $\nu_{m,2}$, $\nu_{m,3}$, $\nu_{m,4}$} with $m$ properly chosen.}
\label{tuning}
\end{figure}

Taking advantage of the customizable spectral mode structure of our resonator design, the large gap of one azimuthal FSR can be bridged using a set of fundamental axial modes. As illustrated in Fig.~\ref{tuning}, tuning of the resonance frequency over the spectral spacing between adjacent axial modes enables resonant insertion of light at any arbitrary frequency. The tuning scheme is implemented by elastically deforming the resonator through mechanical strain, thereby changing its diameter and the refractive index of the medium.

\begin{figure}
\centering
\includegraphics[width=0.45\textwidth]{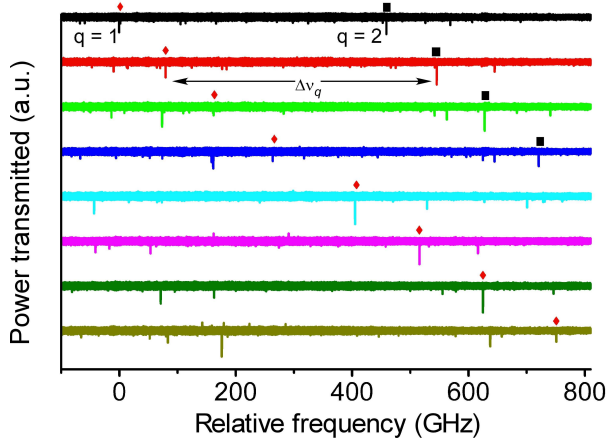}
    \caption{Experimental demonstration of the tuning scheme in Fig.~\ref{tuning}~(b) for the $q = 1$ (diamonds) and $q = 2$ modes (squares) in a bottle microresonator. Mechanically straining the resonator shifts the mode spectrum to higher frequencies. Consecutive spectra (from top to bottom) are recorded with increasing strain, realized by linearly incrementing the voltage applied to the bending piezo which pulls on the end of the resonator fiber. The $q = 1$ mode is shifted across the $q = 2$ mode by a factor of 1.7 times the axial mode spacing $\Delta\nu_{q}$.}
\label{tuning_expt}
\end{figure}

To verify this scheme, we examined the resonator presented in Fig.~\ref{Ringdown_linewidth}~(a), which has a FSR of 2~THz and an axial mode spacing of 425~GHz \cite{Poe09Ult}.  In our setup the fiber ends supporting the resonator are attached to piezoelectric bending actuators offering an elongation of up to 160~$\mu$m. The resonance frequency can thus be shifted by applying a voltage to the actuator which mechanically strains the resonator.  The tuning of two modes with axial quantum numbers $q = 1$ and $q = 2$ over 700~GHz is presented in Fig.~\ref{tuning_expt} and represents a tuning range of 1.7 times the axial mode spacing, or equivalently, 700,000 linewidths. This tuning scheme exceeds all others reported for monolithic microresonators by at least a factor of two and is only limited by the travel range of the bending actuators. The mechanical strain applied to the fiber never exceeds about 0.7 ~GPa, which is 14~\% of the tensile strength of silica fiber \cite{Bar03}. This corresponds to a frequency shift of 0.2~\% of the optical frequency. Repeated tuning of the resonator is therefore possible without damage to the glass.

We note the recent demonstration of strain tuning over 2.2 azimuthal FSRs in a novel ``microbubble resonator'' with a diameter of 200~$\mu$m \cite{Sum10Sup}. This work demonstrated an impressive frequency shift of 0.35~\% of the optical frequency. Currently, UHQ modes have not yet been observed and the small mode volume discussed in the present work has not yet been realized with this resonator. However, ultra-high $Q/V$-ratios should be achievable by correspondingly optimizing the fabrication process.

\section{Optimization of resonator geometry}
\label{sec:Optmize}
For a given in-coupled power, the ratio of $Q/V$ is proportional to the intra-cavity intensity and thus enhances light--matter interactions. For the investigation of CQED systems, the choice of $Q$ and $V$ will influence the behavior of the atom--cavity system (see Refs.~\cite{Vah03,Ilc06}). The single photon Rabi frequency gives the rate of energy transfer between the cavity electric field and the atom, which is double the atom--cavity coupling strength, $\Omega = 2g$. The coupling strength depends on the mode volume according to $g\propto V^{-1/2}$, indicating the cavity should be as small as possible. The observation of strong coupling requires $g$ to dominate the atomic dipole decay rate, $\gamma_{\bot}$, and the cavity field decay rate, $\kappa$, with ($Q\propto1/\kappa$), giving the condition $g\gg(\gamma_{\bot},\kappa)$, where $\gamma_{\bot}=\Gamma/2$ and $\Gamma$ denotes the spontaneous decay rate of the atom. On the other hand, the coupling strength can be related to the saturation photon number, $n_0$, and the critical atom number, $N_0$. The saturation photon number gives the number of photons required to saturate a coupled atom, scaling as $n_0=\gamma_{\bot}^2/2g^2$, while $N_0$, the number of atoms required to have a noticeable effect on the cavity transmission, scales as $N_0=2\gamma_{\bot}\kappa/g^2$. Examining the parameter $n_0$ reveals that $n_0\propto V$, and similarly for $N_0$ we find that $N_0\propto V/Q$. Therefore, the resonator geometry and quality factor govern the scaling of important parameters of the atom--cavity system.

\begin{figure}
\centering
  \includegraphics[width=0.45\textwidth]{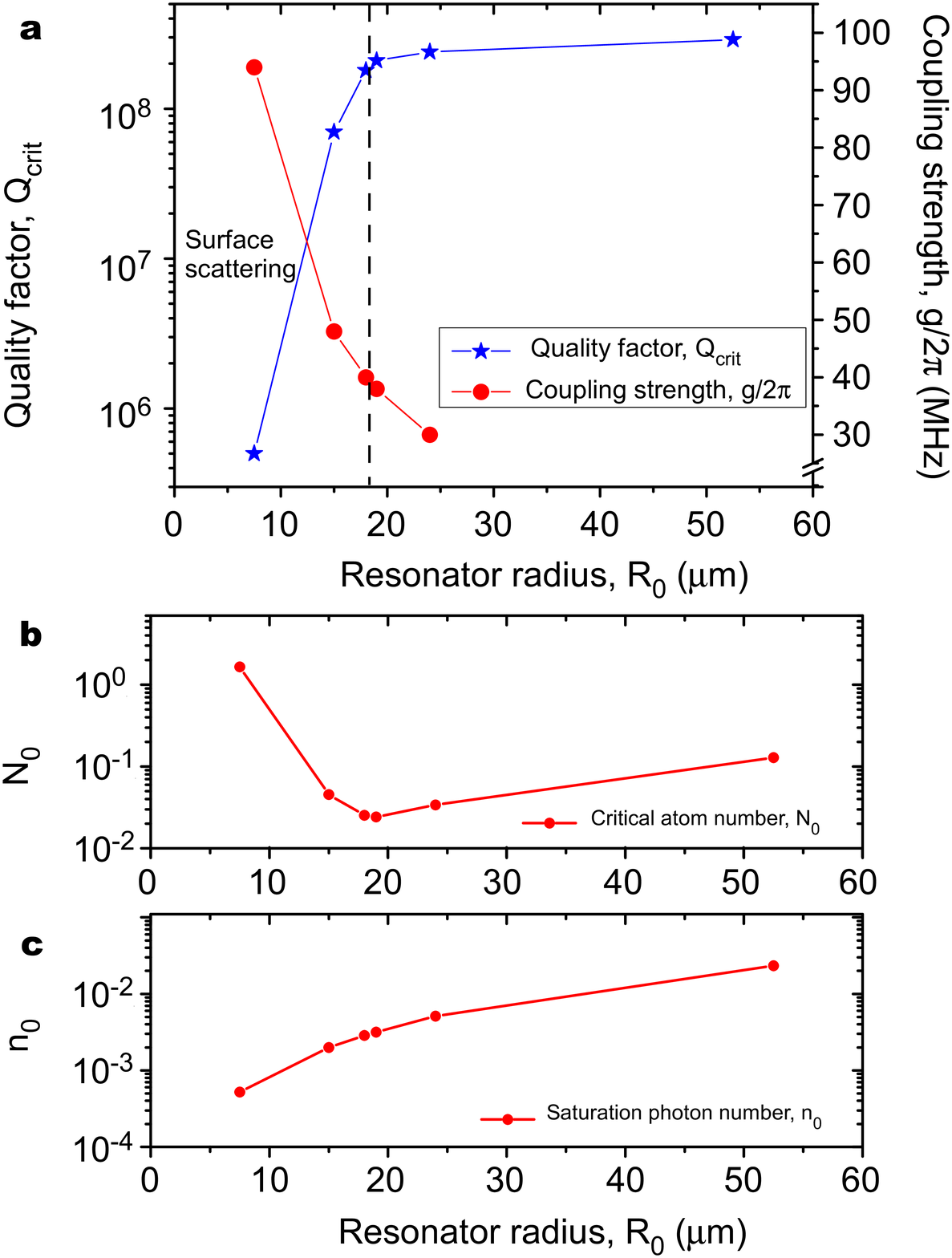}
    \caption{(a) Measured quality factor at critical coupling as a function of resonator radius. The right-hand $y$-axis indicates the expected coupling strength for the D2-transition of a $^{85}$Rb atom with a $q=1$ resonator mode at the resonator surface. Resonators with radii below a critical radius of around 18~$\mu$m experience significant losses, believed to be primarily due to scattering from irregularities on the glass surface. (b) Critical atom number calculated from the data in (a). (c) Saturation photon number calculated from the data in (a). See main text for details.}
    \label{Fig:Q_diameter}
\end{figure}

In order to quantify the performance of our bottle microresonators, we characterized several resonators with radii ranging from 7--53~$\mu$m in terms of quality factor and the predicted coupling strength for the D2 transition of a $^{85}$Rb atom ($\Gamma/2\pi = 6.06$~MHz). Figure~\ref{Fig:Q_diameter}~(a) clearly defines two distinct regions that meet at a critical resonator radius of about 18~$\mu$m. Smaller resonators experience a pronounced quality roll-off of almost three orders of magnitude when the radius is reduced to half the critical radius. This roll-off in quality factor is correlated with a significant increase in the fraction of the field that is evanescent. Consequently, surface scattering and absorption losses from water adsorbed to the surface become more important. Earlier work observed that the quality factor scales with the square root of the resonator radius when losses due to surface scattering and surface water absorption are present, identifying these as dominant loss mechanisms \cite{Ver98}. Furthermore, the mode spectra reveal a sudden onset of large mode splitting for radii below the critical radius. In this case, surface scattering is strong enough to couple light into the anti-clockwise propagating mode, therefore causing the quality factor at critical coupling to decline \cite{Weiss95}. Additionally, for very small resonators, radiative losses also become increasingly important. In stark contrast is the right-hand region where the quality factor remains essentially constant for radii up to 53~$\mu$m.

For the 18~$\mu$m-radius resonator considered in Sec.~\ref{sec:Spectral}, the inferred coupling strength in Fig.~\ref{Fig:Q_diameter}~(a) is found to be high enough to place the atom--cavity system deep into the strong coupling regime. Smaller resonators yield higher $g$ due to the tighter confinement of the resonator mode, but at the expense of quality factor. Plotting the critical atom number for different radii in Fig.~\ref{Fig:Q_diameter}~(b) shows a minimum of $2.4\times10^{-2}$ for a radius of 19~$\mu$m, corresponding to the radius where the $Q/V$ ratio is maximized. The strong quality factor roll-off mentioned above dominates the critical atom number for smaller radii which rapidly approaches unity for a radius of 7~$\mu$m. Larger resonators only show a weak increase in critical atom number which remains a factor of 5 below unity even for a 53-$\mu$m radius resonator. The saturation photon number, shown in Fig.~\ref{Fig:Q_diameter}~(c), can be as low as $5\times 10^{-4}$ for a 7~$\mu$m radius resonator with a corresponding mode volume of 330~$\mu$m$^{-3}$, a coupling strength of 94~MHz, and a moderately high quality factor of 0.5 million. For the 18~$\mu$m radius resonator the saturation photon number is still $3\times 10^{-3}$. Even for the largest resonator the saturation photon number is a factor of 50 below unity.

As mentioned in Sec.~\ref{intro} for nonlinear optics applications, a more relevant quantity to optimize is $Q^2/V$. Taking the data in Fig.~\ref{Fig:Q_diameter}~(a), we find this ratio is as high as $2.1\times10^{13}$~$(\lambda/n)^{-3}$ for a radius of about 19~$\mu$m. This value is among the highest realized for optical microresonators \cite{Kip04Dem,Tan07,Hoo00The}.

Apart from optimizing the resonator radius discussed here and the spectral mode spacing discussed in Sec.~\ref{sec:Spectral}, it is also possible to accurately control the tuning response of the resonator. This issue is particularly relevant when designing the resonator fiber to operate with particular piezo actuators having a fixed travel range. The tunability of resonator modes is governed not only by the resonator itself, but also by the supporting fiber. For a given fiber elongation, $\delta_{\rm tot}$, the applied strain, $P$, and thus the tuning range, is inversly proportional to the length of the support fiber, $L$, and proportional to the support fiber cross-section, $A$. We therefore have $P=\delta_{\rm tot}EA/L$ where $E$ is the Young's modulus of the fiber material. Using a suitable design for the resonator fiber, we are able to exploit this extra degree of customizability using our fiber pulling rig to manufacture complex, non-exponential fiber profiles \cite{Warken08,tesis_Warken}. This feature is particularly important for the CQED experiments described in Sec.~\ref{CQED} due to the limited travel range of the employed piezo actuators.

\section{Frequency Stabilization of UHQ modes in air and vacuum}
\label{sec:Stable}
The modular design of Fabry--Perot resonators enables their full tunability. For example, consider the FP microresonator from Ref.~\cite{Khud08} consisting of two mirrors separated by 160~$\mu$m. Changing the mirror separation by 0.5~$\mu$m using a piezoelectric actuator, results in a resonance frequency shift of 1.2~FSRs (1.1~THz). On the other hand, once the cavity is tuned to the atomic transition, the mirror separation has to be actively stabilized using the Pound--Drever--Hall (PDH) technique. Due to the high finesse of $\mathcal{F}=1\times10^6$, the Fabry--Perot cavity length has to be controlled better than $\delta L = \lambda/(2 \mathcal{F})=0.4 $~pm in order to stabilize the cavities resonance frequency within its linewidth of 0.9~MHz ($Q=3.8\times10^8$).

In contrast, due to their monolithic design, WGM resonators exhibit an excellent passive frequency stability. Therefore, in the pioneering experiments with microtoroidal resonators in Ref.~\cite{Aok06Obs}, it was possible to manually stabilize the resonator frequency via the resonator temperature using a Peltier element and to observe strong coupling of Cs atoms and a mode with $\kappa=2\pi$17.9~MHz ($Q_{\rm crit}= 9.7 \times 10^{6}$). Using a bottle resonator, it should be possible to simplify such experiments by using the tuning scheme presented in Sec.~\ref{sec:Spectral} and by actively stabilizing the UHQ mode frequency. Active stabilization of a bottle mode with $Q_{\rm crit}= 1\times10^8$ to the D2 transition of $^{85}$Rb has been demonstrated both in air and vacuum in our group \cite{Osh10,Osh11}.

\begin{figure}
\centering
  \includegraphics[width=0.45\textwidth]{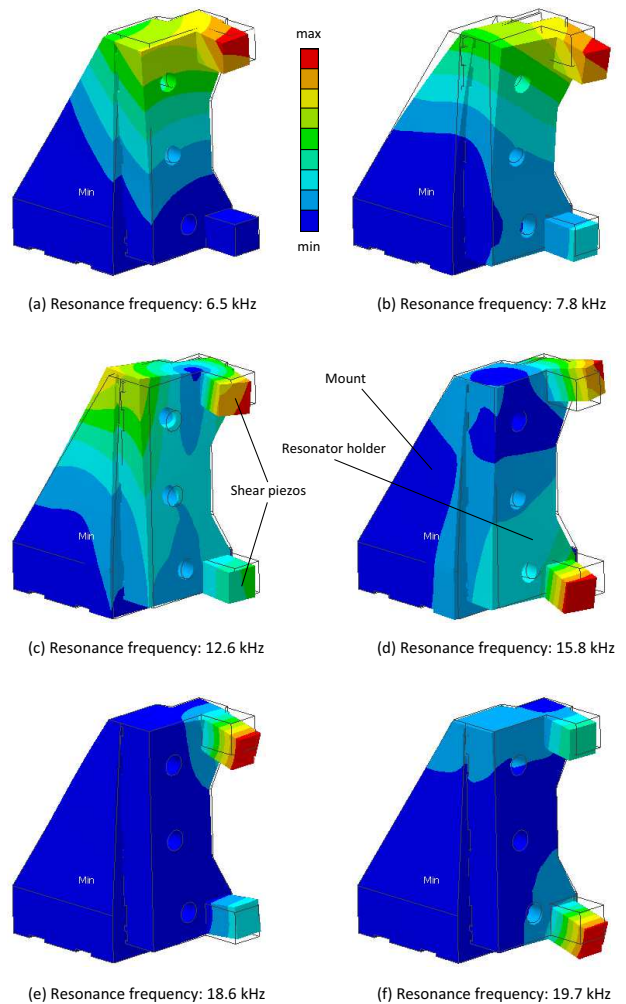}
    \caption{Finite element simulations of the six lowest mechanical resonance frequencies of the resonator holder device. The resonator fiber (not shown) is held between two shear piezos that are glued to a resonator holder which is in turn mechanically attached a triangular mount. The amplitude of the relative motions is indicated by color coding (linear scale) and is also graphically displayed (displacement is exaggerated).}
    \label{Holder_resonance}
\end{figure}

\begin{figure}
\centering
  \includegraphics[width=0.45\textwidth]{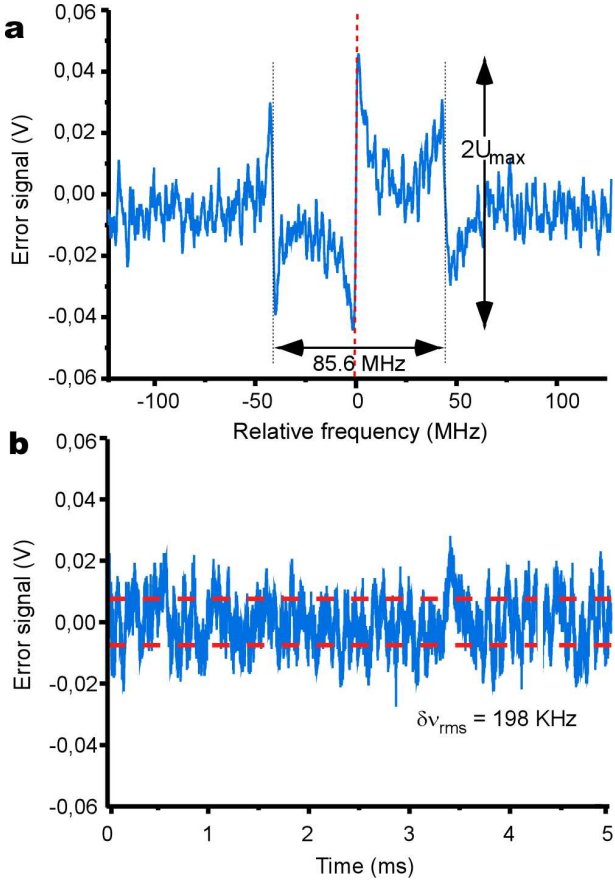}
    \caption{Frequency stabilization of a 50-$\mu$m diameter bottle microresonator in air at 780~nm wavelength. (a) Out-of-lock PDH error signal of a resonance with $Q_{\rm crit} =1.7\times10^8$ measured by scanning the frequency of the resonator mode across a fixed lock-laser frequency. The dotted line at zero relative frequency indicates the center of the resonance to which the resonator is locked in (b). From the noise in the error signal of the stabilized resonator, we calculate an RMS frequency noise of 198~kHz or 9~\% of the resonator linewidth. The bandwidth of the lock is 4~kHz and the optical power in the coupling fiber is 500~nW.}
    \label{PDH_air_locked_combined}
\end{figure}

To achieve active stabilization of UHQ bottle microresonators in our CQED experiment, it was necessary to design a new resonator mount compatible with the requirements of ultra high vacuum (10$^{-10}$~mbar) while maintaining full tunability over 400~GHz and operation with a high bandwidth up to the kHz range. Finite element simulations were critical in ensuring our design was rigid enough to push mechanical resonances up to several kHz. Figure~\ref{Holder_resonance} shows the resonator mount and the associated support structure with the first six resonances occurring primarily around the shear piezos used for tuning and stabilization. By using only the lower piezo for active stabilization and the top piezo to apply a constant offset voltage, we can neglect the first three resonances (Figure~\ref{Holder_resonance}~(a) to (c)) because they will not be excited. This assumption is reasonable since external sources of vibration should be minimal because the mount and coupling setup is carefully isolated from external vibrations using viton cushions in the vacuum chamber. The mechanical motion of the fourth resonance is not in the direction of motion of the piezo and may similarly not be excited. The fifth and sixth resonances (Figure~\ref{Holder_resonance}~(e) and (f)) around 19~kHz show clear oscillations in the direction of piezo motion and set an upper limit on the achievable stabilization bandwidth.
As a demonstration of the stabilization technique, we have stabilized a resonator with an ultra-high quality factor of $Q_{\rm crit} =1.7\times10^8$ in air. The Pound--Drever--Hall error signal shown in Fig.~\ref{PDH_air_locked_combined}~(a) is recorded by scanning the frequency of the resonator mode over the fixed frequency of the stabilization laser light. We then stabilize the center of the resonance to the laser frequency and fluctuations in the error signal around this point are shown in Fig.~\ref{PDH_air_locked_combined}~(b). The locked resonator has an RMS noise of 198~kHz or 9~\% of the resonator linewidth. This RMS noise value is estimated from the noise in the error signal which we convert to resonator frequency noise under the assumption that this is the dominant noise component. In separate measurements, we validated this method with another technique that directly estimates the noise from the transmitted power in the fiber as described in Ref.~\cite{Osh10}. Interestingly, the fractional frequency uncertainty is $5\times10^{-10}$, which, for a $q=1$ axial mode, would be close to the fundamental limit of around $2\times10^{-10}$ imposed by the thermorefractive noise of silica \cite{Mat07}. The actual axial mode number used in this measurement is not known, but optical measurements indicate $q\approx 80$. These measurements have been well reproduced in ultra high vacuum where we have achieved an RMS frequency noise of 344~kHz or 7~\% of the linewidth of a resonator with $Q_{\rm crit} =0.7\times10^8$. Achieving this level of control with thermal stabilization would require the temperature to be stabilized to about 90~$\mu$K, a technically difficult task to accomplish. Recent technical improvements of our photodetector have enabled the lock-laser power in the resonator to be reduced by more than one order of magnitude to 20~nW, thereby reducing light absorption to negligible levels. The stabilization scheme runs stably for several hours without intervention.

\section{Optical switching and memory using the Kerr effect}
\label{sec:Switch}
\subsection{Overview and state-of-the-art}
\begin{figure}
\centering\includegraphics[width=0.2\textwidth]{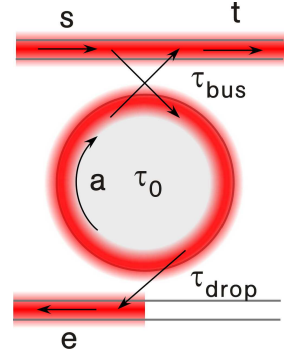}
    \caption{A bottle microresonator, coupled to two ultra-thin fibers in the add--drop configuration. The output field amplitude of the drop fiber is denoted by $e$, the intracavity field amplitude is denoted by $a$, the input field amplitude is denoted by $s$, output field amplitude of the bus fiber is denoted by $t$, and the time constant of the energy transfer between the resonator mode and the drop fiber is $\tau_{\rm drop}$. Similarly, the time constant for the bus fiber is $\tau_{\rm bus}$. The intrinsic lifetime of the resonator is denoted $\tau_0$. The fiber that is used to couple light into the resonator is called bus fiber. The second fiber is called drop fiber. In this configuration, the bottle microresonator acts as a filter which frequency selectively transfers light from the bus fiber to the drop fiber.}
\label{model_adddrop}
\end{figure}

The performance of optical networks and optical communication technology critically depends on devices operating at high speeds and with low power consumption. Optical switches, in particular, are used to switch light between data channels and typically rely on strong light--matter nonlinearities. Microresonators can greatly enhance these nonlinearities because, for a given input power, the nonlinear shift of the resonance frequency in units of the resonator linewidth is proportional to $Q^2/V$. Microresonator-based switches considered here are based on the add--drop filter design where light can be routed between two different output ports as shown in Fig.~\ref{model_adddrop} or, in the particular case of the bottle resonator, in Fig.~\ref{bottle_scheme}~(b).

The only high efficiency add--drop filters (less than 1~\% insertion loss, power transfer efficiencies exceeding 90~\%) based on ultra-high quality factor microresonators so far have been realized with WGMs \cite{Poe10,Rok04Ult}. Here, we give an overview of the performance of state-of-the-art all-optical switches based on various microresonator designs and characterize them. Not all experiments cited here are aimed at achieving the lowest possible switching threshold. Often, a compromise between the switching threshold, $P_{\rm th}$, and the bandwidth, $B$, is chosen. We therefore consider the constant ratio $P_{\rm th}/B^2$ as a figure of merit, thereby ensuring a fair comparison between different systems \cite{tesis_Poellinger}. The characteristics of several all-optical switches are summarized in Table~\ref{tab:switch}. Optical switching can, amongst others, be achieved using free carrier nonlinearities induced by one- or two-photon absorption in semiconductors. However, many of the corresponding experiments rely on two-wavelength pump-probe optical switching schemes and will not be discussed here further (see for example Ref.~\cite{Alm04All}).

The lowest bistable switching thresholds were observed in a photonic crystal microresonator made of GaAs with a threshold of $P_{\rm th} = 6.5$~$\mu$W  \cite{Not05Opt,Wei07Non}. The bandwidth of such a thermo-optical switch is, however, only on the order of 1~MHz because the thermal relaxation times in photonic crystal cavities amount to at least 100 ns \cite{Not05Opt,Har09Ext}. These switches are thus fundamentally limited in their switching speed and yield a value of $P_{\rm th}/B^2=6.5$~$\mu$W/MHz$^2$. Moreover, to the best of our knowledge, no add--drop functionality has been experimentally realized with photonic crystal cavities so far. This limits their use in all-optical signal processing to ``ON-OFF'' switching of the power in one channel while switching of signals between two channels is not possible. Thermo-optical bistability has also been observed in silicon ring-resonators at a switching threshold of 1.3~mW \cite{Alm04Opt}. Here, the bandwidth was found to be 500~kHz, which results in a $P_{\rm th}/B^2$-ratio of 5200~$\mu$W/MHz$^2$, three orders of magnitude worse than for photonic crystals.

Free carrier nonlinearities were employed to realize single-wavelength optical switching in bistable semiconductor etalons, see Ref.~\cite{Pey85Opt} and references therein. In these devices, typical switching threshold powers exceed 1~mW and the bandwidth is around 100~MHz, limited by the carrier recombination time and/or diffusion speed. This enables ON-OFF switching with an extremely low value of $P_{\rm th}/B^2= 0.3$~$\mu$W/MHz$^2$. Operation in transmission mode, necessary for the operation as an add--drop filter, has not yet been demonstrated with bistable etalons. One technical challenge in this context is their relatively high intrinsic loss. Note also that the bistable etalons need to be optimized for operation at a particular wavelength due to the strongly wavelength-dependence of the free carrier nonlinearities. The Kerr effect, on the other hand, prevails over a much larger spectral range.

Due to the large $Q/V$ ratios found in bottle microresonators, the observed $P_{\rm th}/B^2= 4.5$~$\mu$W/MHz$^2$ ranks among the lowest to date \cite{Poe10}. In addition, they are suitable for both ON-OFF switching and as add--drop filters. Such a system is theoretically described in the next section. More details on the experimental setup can be found in Ref.~\cite{tesis_Poellinger}.

\begin{table}
\centering
\begin{tabular}{llll}
\hline\noalign{\smallskip}
 System   &  Add/Drop  &$P_{\rm th}/B^2$                & Ref.\\
          &            &   ($\mu$W/MHz$^2$)  &             \\
\noalign{\smallskip}\hline\noalign{\smallskip}
  Silicon etalon           & No  & 0.3 & \cite{Pey85Opt}  \\
  WGM Bottle               & Yes & 4.5 &  \cite{Poe10}   \\
  Photonic Crystal/         &     &     &                 \\
  GaAs                     & No  & 6.5 & \cite{Not05Opt,Wei07Non}\\
  Silicon ring             & No  & 5200& \cite{Alm04Opt} \\
\noalign{\smallskip}\hline
\end{tabular}
\caption{Comparison of state-of-the-art all-optical switches. Only the bottle microresonator has been operated as an add--drop filter.}
\label{tab:switch}
\end{table}

\subsection{Model for Kerr bistability in bottle resonators}
\label{Modellkerrswitching}
Kerr bistability in a bottle microresonator in add--drop configuration causes a shift in the resonance frequency $\Delta\omega_0(I)$ depending on the intracavity intensity $I(t)$. This system is schematically shown in Fig.\ref{model_adddrop} and can be described by the differential equation for the intracavity field, $a(t)$, \cite{Hau84Wav}
\begin{equation}
\begin{split}
\frac{d}{dt}a(t) ={} &i(\omega_0 + \Delta\omega_0(I)-\omega) a(t)\\
                     &-\frac{1}{2}\left(\tau_0^{-1} + \tau_{\rm bus}^{-1} + \tau_{\rm drop}^{-1}\right)a(t)\\
                     &+ \tau_{\rm bus}^{-1/2} s(t)~.
\end{split}
\label{Eqn coupling 1mod}
\end{equation}
In the following, the intracavity intensity is modeled as a boxcar function of the radial coordinate within the FWHM radii $r_1$ and $r_2$ of the normalized radial intensity distribution $I(r)/I_{\rm max}$ of the fundamental mode in a 35-$\mu$m diameter resonator. The peak value of the intracavity intensity is \cite{Hecht}
\begin{equation}
\label{int_dist_model}
\begin{split}
I(t)&= \frac{\lvert a(t) \rvert^2 cn^2}{V} \cdot \frac{\int_{r_1}^{r_2} I(r)/I_{\rm max}\cdot rdr}{\int_{r_1}^{r_2}rdr}\\
    &= 0.81 \frac{\lvert a(t) \rvert^2 cn^2}{ V}~.
\end{split}
\end{equation}
The nonlinear resonance frequency shift can thus be written as
\begin{equation}
\begin{split}
\Delta\omega_0(t) &= -\omega_0 \frac{n_2 I(t)}{n}\\
&=-\frac{1.62 \cdot \pi n\cdot n_2 c^2 }{\lambda_0 V} \lvert a(t)\rvert^2\\
&\equiv -c_{\rm Kerr} \lvert a(t)\rvert^2~,
\end{split}
\end{equation}
with a nonlinear refractive index of silica given by $n_2 = 2.5\times 10^{-20} $~W/m$^2$ at a wavelength of $\lambda_0=852$~nm \cite{Lee02}. For a given ``cold'' laser--resonator detuning $\delta_0 = \omega-\omega_0$, i.e., the laser--resonator detuning for vanishing input powers, one obtains
\begin{equation}
\begin{split}
\frac{d}{dt}a(t)={} &-i\left(c_{\rm Kerr}\lvert a(t)\rvert^2 + \delta_0 \right) a(t)\\
                    &-\frac{1}{2}\left(\tau_0^{-1} + \tau_{\rm bus}^{-1}+ \tau_{\rm drop}^{-1}\right)a(t)\\
                    &+ \tau_{\rm bus}^{-1/2} s(t)~.
\end{split}
\label{master1}
\end{equation}
In order to model the experimentally observed bistable behavior presented below, the input wave amplitude is assumed to be
\begin{equation}
\begin{split}
s\left(t\right) &= \sqrt{P_{\rm max}}\sin\left(\pi t/T\right)~, \quad\text{for 0 $\leq$ t $\leq$ T} \\
s\left(t\right) &= 0~, \qquad\qquad\qquad\qquad\text{for t $<$ 0 and t $>$ T}~.
\label{master2}
\end{split}
\end{equation}
The optical power incident through the bus fiber, $P_{\rm in} = |s|^2$, thus corresponds to a $\sin^2$-shaped pulse with a FWHM pulse duration of $\tau_{\rm pulse}= T/2$.
For critical coupling, i.e., $\tau_{\rm bus}^{-1}=\tau_0^{-1} + \tau_{\rm drop}^{-1}$, the time constants can be expressed using the loaded quality factor, $Q_{\rm load}^{-1}= 1/\left(\omega_0\tau_{\rm 0}\right)  +1/\left(\omega_0 \tau_{\rm bus}\right)  +1/\left(\omega_0\tau_{\rm drop}\right)$, and the intrinsic quality factor $Q_0$
\begin{equation}
\begin{split}
\tau_0^{-1} &=\frac{\omega_0}{Q_0}~,\\
\tau_{\rm drop}^{-1} &= \frac{\omega_0}{2Q_{\rm load}}-\frac{\omega_0}{Q_0}~,  \\
\tau_{\rm bus}^{-1}&= \frac{\omega_0}{2Q_{\rm load}}~.
\end{split}
\end{equation}
We now solve Equation~\ref{master1} numerically for $a(t)$ with $0\leq t \leq T$ using the boundary condition $a(0)=0$.
The relevant quantities are inferred from
\begin{equation}
\begin{split}
P_{\rm in}\left(t\right) &= \lvert s\left(t\right)\rvert^2~,\\
P_{\rm out}^{\rm bus}\left(t\right)&= \lvert -s\left(t\right) +\tau_{\rm bus}^{-1/2}a\left(t\right) \rvert^2~,\\
P_{\rm out}^{\rm drop}\left(t\right)&= \tau_{\rm drop}^{-1} \lvert a\left(t\right)\rvert^2~,\\
\delta\left(t\right) &= \delta_0 + c_{\rm Kerr} \lvert a\left(t\right)\rvert^2~.
\end{split}
\end{equation}
Here, $\delta=\delta_0-\Delta\omega_0$ is the instantaneous laser--resonator detuning when the pulse is applied. Note that  $P_{\rm out}^{\rm drop}$ is proportional to the intracavity energy.

\begin{figure}[]
\centering\includegraphics[width=0.45\textwidth]{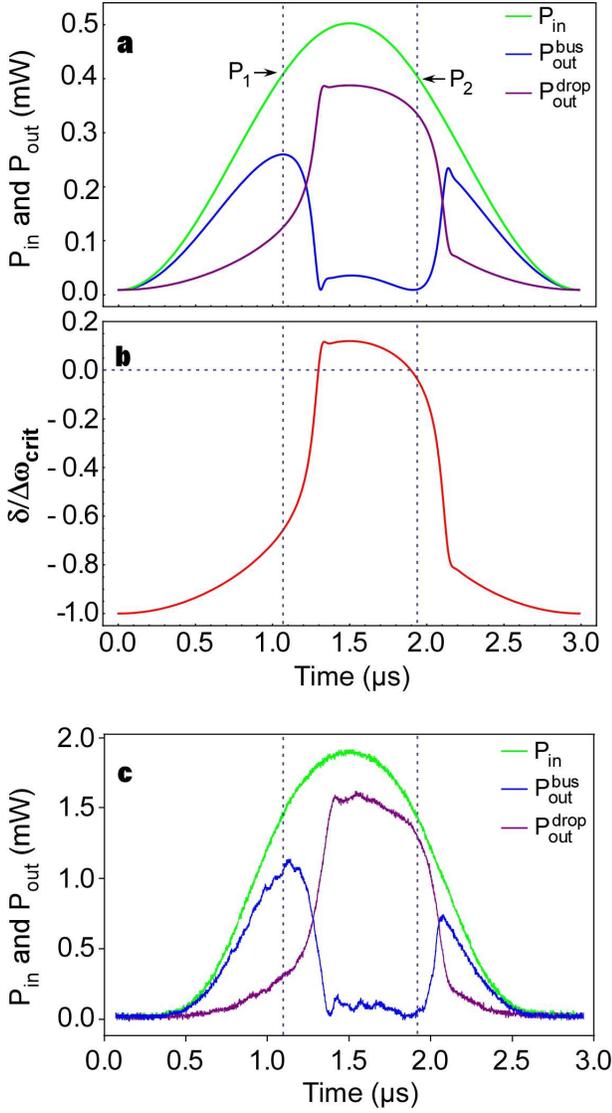}
    \caption{(a) $P_{\rm out}^{\rm bus}$ and $P_{\rm out}^{\rm drop}$ modeled for a $\sin^2$-shaped pulse of the input power $P_{\rm in}$. A resonator with parameters $Q_0 = 1.8\times 10^8$, $Q_{\rm load} = 1.7\times 10^{7}$ and $V=1000$~$\mu$m$^3$ is assumed. (b) Evolution of the laser--resonator detuning $\delta$. The cold laser--resonator detuning at $t=0$ has a value of $\delta_0=-\Delta\omega_{\rm crit}$. When $P_{\rm in}$ exceeds a threshold $P_{1}$, $d \delta/d P_{\rm in}$ reaches a critical value and the Kerr effect ``pulls'' the resonator mode into resonance with the laser frequency. At the peak value of $P_{\rm in}$, the detuning reaches positive values. When $P_{\rm in}$ decreases again, at a power $P_{2}$, the intracavity intensity ($\propto P_{\rm out}^{\rm drop}$) has sufficiently decreased to reach zero detuning. From this point onward, the resonator rapidly returns to the initial situation. We define the threshold powers $P_1$ and $P_2$ by the condition $d P_{\rm out}^{\rm bus}/d P_{\rm in}=0$. Between the two switching events, the transmission through the bus fiber is strongly reduced due to critical coupling. At the same time, due to the near-zero values of $\delta$, the intracavity power is high and 84~\% of $P_{\rm in}$ is transferred to the drop fiber output. (c) Experimentally measured response of the system to a $\sin^2$-shaped pulse (green). As soon as $P_{\rm in}$ exceeds a certain threshold, the light is resonantly switched to the drop fiber via the Kerr effect.}
    \label{modelpowers}
\end{figure}

\begin{figure}[]
\centering\includegraphics[width=0.43\textwidth]{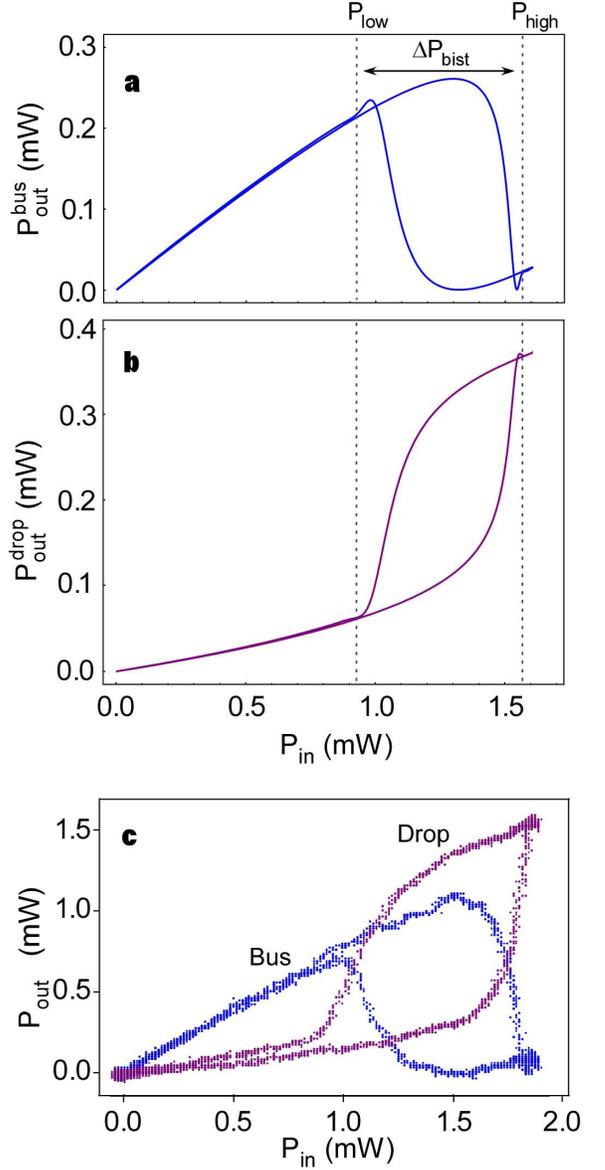}
    \caption{(a) $P_{\rm out}^{\rm bus}$ and (b) $P_{\rm out}^{\rm drop}$ as a function of $P_{\rm in}$. The theoretical model predicts a pronounced hysteretic behavior which, for a certain range of $P_{\rm in}$, exhibits two stable states. The bistable regime is defined as the region between the powers $P_{\rm low}$ and $P_{\rm high}$, where both, $P_{\rm out}^{\rm bus}$ and $P_{\rm out}^{\rm drop}$, can take two different values depending on the sign of $d P_{\rm in}/dt$. Note that the threshold powers from Fig.~\ref{modelpowers}, $P_1$ and $P_2$, do not coincide with the borders of the bistable region defined here. (c) By plotting $P_{\rm out}^{\rm bus}$ and $P_{\rm out}^{\rm drop}$ versus $P_{\rm in}$ for the data in Fig.~\ref{modelpowers}~(c), bistable behavior is apparent and ranges from $P_{\rm low} \approx 1.0$~mW to $P_{\rm high} \approx 1.8$~mW. }
\label{modelbsit}
\end{figure}

\subsection{Experimental observation of optical bistability}
\label{experimentelle Ergebnisse AOS}
In order to verify the findings from the above model, we apply it to data taken in an earlier experiment where a $q=2$ mode with an intrinsic quality factor of $1.8\times 10^8$ and a mode volume of $V=1000$~$\mu$m$^3$ is used in an add--drop configuration \cite{Poe10}. The gap between the resonator and the drop fiber is chosen to yield a loaded quality factor of $Q_{\rm load}= 1.7\times 10^7$ at critical coupling. We choose an initial negative detuning of $\delta_0= \omega_{\rm probe} - \omega_{\rm 0} = -1.2\times\Delta\omega_{\rm crit} $. Under these conditions, 85~\% of the optical power exits the bus fiber on resonance. Figure~\ref{modelpowers}~(c) shows the experimentally observed powers at the bus and drop fiber waists for a $\sin^2$-shaped modulation of $P_{\rm in}$ with a FWHM pulse duration of $\tau_{\rm pulse} = 1.25$~$\mu$s. In the simulation of Fig.~\ref{modelpowers}~(a,b), a cold laser--resonator detuning of one linewidth $\delta_0=-\Delta\omega_{\rm crit}$, a pulse duration of $\tau_{\rm pulse}= 1.25$~$\mu$s, and a peak power $P_{\rm max}= 0.51$~mW are assumed. For the given parameters, the value of $P_{\rm max}$ was chosen by trial and error. It was stepwise increased until $\delta$ reached positive near-zero values in the central region of the input pulse.

In order to clearly show the bistable behavior, we also plot $P_{\rm out}^{\rm bus}$ and $P_{\rm out}^{\rm drop}$ as a function of $P_{\rm in}$. The result of the simulation is shown in Fig.~\ref{modelbsit}~(a,b) and is in good qualitative agreement with the measurement in Fig.~\ref{modelbsit}~(c).

Note that the model presented in Sec.~\ref{Modellkerrswitching} can only be qualitatively compared to the experimental data because it is based on an over-simplified description of the modal intensity distribution, cf. Eq.~\ref{int_dist_model}. For this reason, the frequency shift induced by the Kerr effect for a given incident power is not described correctly at a quantitative level. At the qualitative level, however, we find good agreement both in terms of the threshold power $P_1$ and of the characteristic evolution of the bus and drop fiber transmissions during the pulse. The modeled threshold power is only a factor of three smaller than the measured value. The good qualitative agreement of the above measurements with the predictions from the model strongly suggests that that the Kerr effect is at the origin of the bistable behavior observed on this time scale. Furthermore, pulsed experiments in Ref.~\cite{Poe10} show no dependency between the bistability threshold and pulse duration as long as the latter is much shorter than the thermal response time (13--15~ms) of the mode volume. This also indicates that the Kerr effect is the prevailing non-linear mechanism under the experimental conditions considered here.

\section{All-optical signal processing in bottle microresonators}
\label{sec:All_optical}
We now demonstrate all-optical signal processing functionalities based on the bistable behavior characterized above. As has been shown in the previous section, routing of the signal beam between the bus or the drop fiber exit can be achieved by setting $P_{\rm in}$ to a value below or above the bistable regime, respectively. Furthermore, the optical bistability also allows us to realize an optical memory: For a value of $P_{\rm in}$ chosen within the bistable regime, two stable states at the outputs are possible and the actual state depends on the history of the system.

\begin{figure}
\centering\includegraphics[width=0.45\textwidth]{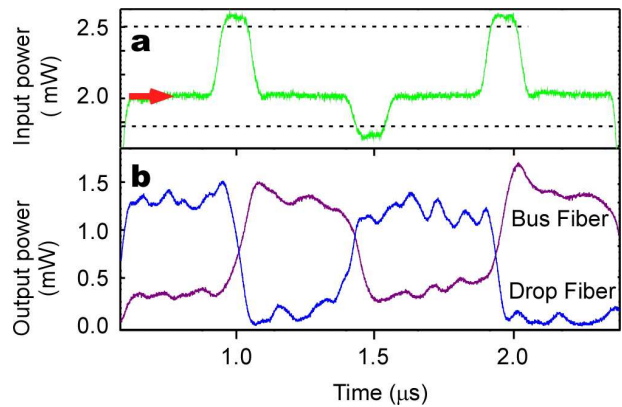}
    \caption{Demonstration of optical memory functionality using the  Kerr effect in a bottle microresonator in add--drop configuration. (a) For an input power level in the bistable regime, $P_{\rm bist}$, as indicated by the arrow, the power at the output ports of the bus fiber and of the drop fiber exhibits two stable states, see (b). The output state is chosen by temporarily lowering (raising) $P_{\rm in}$ below (above) the bistable regime, indicated by the dashed lines in (a).}
\label{optical_memory}
\end{figure}

In order to realize such an optical memory in the experiment, we choose an operating power centered in the bistable region, $P_{\rm in} = P_{\rm bist} \approx 2$~mW, see Fig.~\ref{optical_memory}. Changing $P_{\rm in}$ to a power higher than $P_{\rm high}$ or lower than $P_{\rm low}$ switches between the two possible output states. When returning to $P_{\rm bist}$, the system then stays in the chosen state.

Single-wavelength optical memories have so far been realized with microring resonators using thermally induced bistability \cite{Alm04Opt} and with etalons using free carrier nonlinearities \cite{He93All}. In both cases, only ON-OFF functionality was demonstrated since no drop channel was implemented. In contrast, the optical memory realized here is a true add--drop device, meaning that it allows one to route a signal between its two output ports. Since the optical memory is operated in the bistable regime, it is quite sensitive to thermal fluctuations and drifts. Therefore, the residual absorption of the signal light field currently limits the storage time of this optical memory to the sub-microsecond range. A similar limiting effect has also been reported in Ref.~\cite{He93All}.

\section{Coupling spatially separated WGM resonators}
\label{sec:Separate}
The coupling and interaction of resonators forming a network is of fundamental and practical interest. Coupled WGM resonators are starting to receive considerable interest because of their ability to produce electro-magnetically-induced transparency-like (EIT) phenomena (see Ref.~\cite{Xiao08}, and references therein). These systems are particularly advantageous since they can, in principle, produce narrow EIT-like windows at any wavelength within the material transparency window, and may be useful for optical delay lines by tuning the group velocity of resonant photons. As another example, a recent proposal investigated a strongly interacting many-body system composed of an array of resonators, e.g., WGM resonators, and interacting polaritons, which raised the prospect of a WGM-resonator-based quantum simulator \cite{Har06}. Finally, the development of mechanisms to enable the interaction of remote qubits over a CQED quantum network is a major technical goal that remains an open issue \cite{Kim08}. In particular, it requires classical strong coupling between remote resonators, i.e., the ability to transfer the field stored in one CQED-compatible resonator to a second, remote resonator and back on a timescale much shorter than the photon lifetime inside the resonators. In the following, we present a first study that examines the feasibility of observing classical strong coupling between two bottle microresonators.

\subsection{Coupling between modes of different polarization in a single microresonator}
\begin{figure}[]
\centering\includegraphics[width=0.45\textwidth]{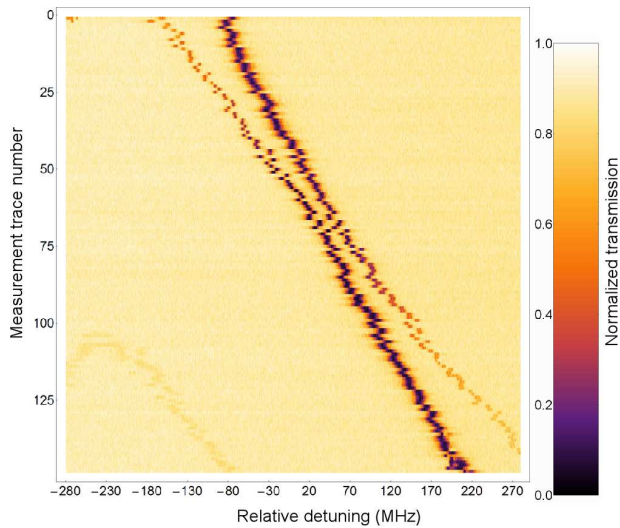}
    \caption{Avoided crossing between two modes of different polarization in a single resonator. Increasing numbers on the $y$-axis correspond to increasing mechanical strain applied to the resonator. The different tuning rates of modes having different polarization make it possible to change the relative detuning between the two modes from negative to positive. When the relative detuning becomes comparable to the coupling rate between the modes, $2g_{\rm res}$, they loose their individual character and the new eigenmodes exhibit a minimum frequency splitting equal to $2g_{\rm res}$.}
\label{Polarisationskoppllplot}
\end{figure}

As an initial step, we examine the possibility of coupling two different WGM modes of the same resonator which, in particular, are orthogonally polarized. When strain tuning a bottle microresonator, WGM modes of TE and TM character exhibit different tuning speeds due to the anisotropy of the strain-induced refractive index change. More quantitatively, the tuning rate differs by a factor of around 1.6, with the TE mode tuning faster since its electric field is perpendicular to the resonator surface. Coupling between TE and TM modes can be mediated by scattering as has been demonstrated with microspheres in Ref.~\cite{vonKlitzing01} (see also Ref.~\cite{Weiss95}). Our measurements show a coupling rate of up to $2g_{\rm res}=25$~MHz as determined from the width of the avoided crossing, see Fig.~\ref{Polarisationskoppllplot}. These results confirm the possibility of strongly coupling two WGM modes within a single resonator. The next obvious goal is to strongly couple the modes of two remote resonators.

\subsection{Transmissive and reflective coupling of two microresonators}
\begin{figure}[]
\centering\includegraphics[width=0.45\textwidth]{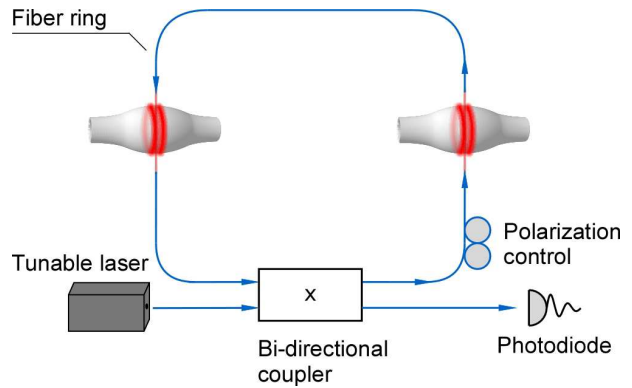}
    \caption{Setup for transmittive coupling between two bottle microresonators via a fiber ring resonator.}
\label{ring_duores}
\end{figure}

\begin{figure}[]
\centering
\includegraphics[width=0.45\textwidth]{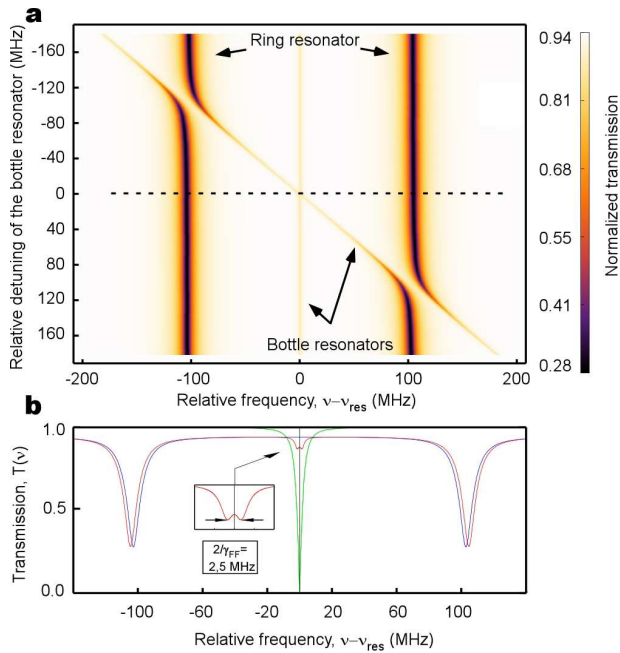}
    \caption{Numerically calculated transmission of a fiber ring resonator with a finesse of 24 and a free spectral range of 210~MHz, critically coupled to two coupled bottle microresonators with $Q=10^8$. (a) The $x$-axis shows the detuning of the laser relative to the first bottle microresonator while the $y$-axis shows the relative detuning of the second bottle microresonator with respect to the first one. The resonance frequency of the first bottle microresonator is assumed to be centered between two consecutive resonances of the fiber ring resonator. The coupling strength between the bottle microresonators and the fiber ring is set to 20~MHz, leading to a large avoided crossing when the second bottle microresonator is tuned into resonance with the fiber ring resonator mode. A cross-section along the dotted line, corresponding to the case where the two bottle modes are resonant, is shown in (b). In this case, the coupling between the bottle modes leads to a small but visible avoided crossing. From this plot, the coupling strength between the bottle modes is inferred to be 2.5~MHz. For comparison, the bare transmission functions of the bottle microresonators and of the fiber ring resonator are also shown (green and blue lines).}
\label{Fiber_loop_simul}
\end{figure}

Coupling between modes of two remote resonators can be achieved by two complementary methods; namely, transmissive and reflective coupling. In the former case, we couple two clockwise propagating modes via an optical fiber loop with two ultra-thin sections which act as fiber couplers. The out-coupled light of the first resonator propagates along the fiber loop and excites the mode of the second resonator. It is then out-coupled again and looped back to the first resonator. The setup is schematically shown in Fig.~\ref{ring_duores}. In this case, the fiber loop itself acts as a ring resonator and the bi-directional coupler in Fig.~\ref{ring_duores} allows us to couple light into and out of this fiber ring resonator. In order to theoretically investigate this scheme, we simulate it in the situation where the first bottle mode is non-resonant with the fiber ring resonances, as shown in Fig.~\ref{Fiber_loop_simul} \cite{tesis_Andi}. The results show the expected avoided crossings between the bottle modes as well as between the bottle mode and the fiber ring modes. However, the predicted avoided crossing of 2.5~MHz between the bottle modes is comparable to their linewidth. Moreover, the bottle modes only show up as 5~\% transmission dips in the fiber ring resonator spectrum. This makes the observation of the effect a challenging task which we have so far not been able to achieve. Avoided crossing has however been experimentally observed between the second bottle microresonator and fiber ring resonator. As an alternative to a spectral measurement of an avoided crossing in the steady state regime, another promising route to find a signature of the remotely coupled modes might be to measure the transient response of the system \cite{Fer10}.

\begin{figure}[]
\centering\includegraphics[width=0.45\textwidth]{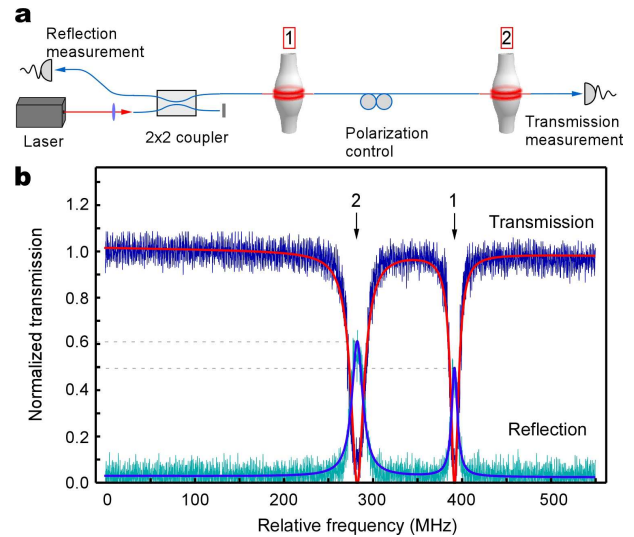}
    \caption{(a) Experimental setup for reflective coupling between two bottle microresonators. (b) Transmission and reflection spectra of critically coupled resonator modes. The loaded quality factors are $Q_1 = 3\times10^7$ and $Q_2=2\times10^7$.}
\label{reflcpl}
\end{figure}

As an alternative, the reflective coupling scheme relies on the backscattering of light from the clockwise into the anti-clockwise propagating bottle modes in each bottle resonator. This backscattering is mediated by surface roughness and, if two resonators are coupled with a straight fiber with two ultra-thin sections, can lead to the following effect: the light is first coupled from the clockwise to the anti-clockwise mode in the first resonator. Then it is out-coupled, propagates along the straight fiber and excites the anti-clockwise mode of the second resonator. From there, it is coupled to the clockwise mode, is out-coupled and propagates back to the first resonator. To our knowledge, this scheme has thus far not been demonstrated. The setup is schematically shown in Figure~\ref{reflcpl}~(a). Note that, in this case, the bottle microresonators act as reflective elements and, in conjunction with the straight fiber, realize a Fabry--Perot-like resonator with a free spectral range fixed by the length of the straight fiber. A similar WGM system has been theoretically modeled in Ref.~\cite{Xiao08}. The experimental spectra in Fig.~\ref{reflcpl} prove that the effect indeed causes back reflection and that 50--60~\% of the light which is critically coupled into the bottle microresonator is detected in the backward direction photodiode. Correcting for transmission losses from the $2\times2$-coupler, splices, and losses in the taper sections we infer the reflection coefficients of the first and second resonator to be $R_1=53\%$ and $R_2=76\%$, respectively. From these values we estimate that the light would on average only undergo about three full round-trips between the first and the second bottle microresonator. There are two methods of increasing $R_1$ and $R_2$: one can either increase the coupling between the clockwise and anti-clockwise modes in each bottle microresonator, or one can increase their intrinsic quality factor and thus reduce the intra-bottle microresonator losses. The first method, however, comes at the expense of reducing the loaded quality factors of the bottle microresonators and is thus not appropriate for CQED applications, cf. Sec.~\ref{sec:Optmize}. Increasing the intrinsic quality factor, on the other hand, is experimentally difficult given that we have already optimized our fabrication process. Summarizing, for implementing the reflective coupling scheme presented here, one requires WGM resonators with a high intrinsic quality factor and, at the same time, a large mode splitting in units of the cavity linewidth. These conditions are fulfilled, e.g., in the case of microtoroidal resonators \cite{Kip04Dem}.

\section{A cold-atom apparatus for CQED experiments}
\label{CQED}
We give an overview of our apparatus used to deliver cold rubidium ($^{85}Rb$) atoms to the evanescent field of the bottle microresonator. We demonstrate its performance in terms of the temperature of the atom cloud near the resonator and the ability to precisely control the height of the turning point of the atom cloud during its parabolic trajectory to the resonator. Finally, we describe our detection scheme designed to observe atom transits through the resonator field.

\subsection{Overview}
Atom trapping in a FP cavities is now a well-understood process and has been achieved in numerous experiments \cite{Kim08,Khud08,Chap04}. The long interaction time between the trapped atom and the light in the cavity allows one to perform complex operations such as feedback cooling and quantum control. As will be discussed in Sec.~\ref{Placing_atom}, this level of control is not yet possible with WGM resonators and experiments must be performed during the brief and random atom transits through the resonator field. The interaction time should ideally be limited by the thermal velocity of the atom and as few atoms as possible should be adsorbed on the resonator surface over time in order to preserve the resonator quality factor. Two options exist: a single vacuum chamber setup in which atoms are dropped onto the resonator from a background-gas loaded 3D-magneto-optical trap (MOT), or a differentially pumped two-chamber setup with the resonator chamber containing a low background pressure and the second chamber containing a high rubidium background pressure. Concerning adsorption of rubidium on the resonator surface, the latter design is preferable.  In this case, several schemes for delivering atoms from the second chamber to the resonator exist: loading a 3D-MOT in the resonator chamber from a 2D$^+$-MOT in the second chamber and then dropping the atoms on the resonator, or loading a 3D-MOT in the second chamber, transporting the atoms into the resonator chamber using an optical conveyor belt \cite{Khud08}, and then dropping the atoms, or finally, an atomic fountain might be used in order to launch atoms on a parabolic trajectory towards the resonator. Dropping the atoms on the resonator, however, suffers from a fundamental limitation. Assuming the atoms are dropped from a height, $h$, of more than 160~$\mu$m, the velocity acquired by the falling atoms, $v_{\rm grav}=\sqrt{2gh}$, will dominate the average thermal velocity of the atoms in the direction of the resonator, $v_{\rm therm}=\sqrt{\frac{9\pi k_BT}{8 m_{\rm Rb}}}$, where $g$ is the gravitational constant, $k_B$ is Boltzmann's constant, $m_{\rm Rb}$ is the atomic mass of rubidium, and $T$ is the atomic temperature, assumed to be 10~$\mu$K \cite{Met99}. Alternatively, an atomic fountain can launch a sub-Doppler cooled atomic cloud such that the heights of the turning points of the atomic trajectories on average coincide with the position of the resonator, thereby yielding the longest atom--field interaction time. Furthermore, it is possible to truncate the vertical velocity distribution by temporal or spatial filtering before the atoms reach the resonator, thereby preventing the unwanted fast atoms from coating the resonator surface. For these reasons, we chose an atomic fountain as the means of delivering the atoms.

\begin{figure}[]
\centering\includegraphics[width=0.25\textwidth]{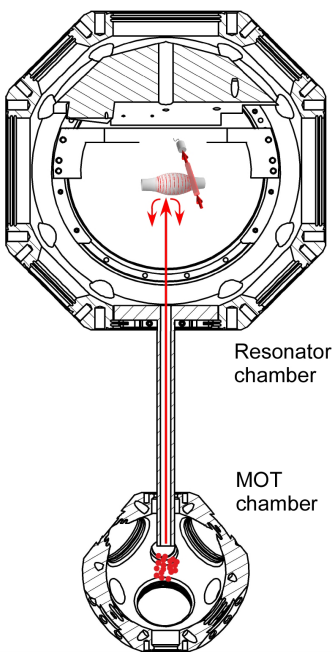}
    \caption{Cross-section through the vacuum apparatus. Rubidium is cooled in a magneto-optical trap in the lower chamber. The atomic cloud is then launched into the resonator chamber using an atomic fountain.}
\label{chamber2}
\end{figure}

A schematic of the fountain is shown in Fig.~\ref{chamber2}. A Rb dispenser in the lower chamber provides a source of hot $^{85}{\rm Rb}$ atoms that are trapped and cooled in a six-beam MOT. The cooling laser beams are arranged in a 1--1--1 configuration with three beams pointing downward at 57.3$^{\circ}$ and three opposing beams pointing upward also at 57.3$^{\circ}$. The magnetic field of the MOT is then switched off and the atoms are transferred into an optical molasses with a vertical velocity, $v_z$, set by the detuning, $\Delta\omega$, between the upward and downward pointing sets of beams. In this moving molasses, the atom temperature is further reduced to about 6--10~$\mu$K using polarization gradient cooling. The amount of detuning governs the velocity through the relation $v_z = \Delta\omega\sqrt{3}/k$, where $k$ is the wavenumber of the cooling light. Differential pumping between the two chambers reduces the background pressure in the resonator chamber by more than a factor of one hundred with respect to the MOT chamber. The base pressure of the MOT chamber without Rb vapor is around $5\times10^{-10}$~mbar and the pressure in the resonator chamber is around $1\times10^{-9}$~mbar without baking. More experimental details can be found in Ref.~\cite{tesis_Hauswald}.

The resonator set-up, consisting of a resonator holder shown earlier in Fig.~\ref{Holder_resonance} as well as two coupling fibers as shown in Fig.~\ref{bottle_scheme}~(b), is compactly mounted upside down on a gold-plated copper block. Its orientation is chosen to yield the best possible access for the atomic fountain while not hindering other optical axes. Alignment of the coupling fibers is performed using UHV-compatible slip-stick piezo translation stages with 50-nm stepping resolution (Attocube systems AG). A light sheet is used to probe the fountain a few centimeters below the resonator, and two zinc selenide viewports provide access for a CO$_2$ laser beam which can be focussed on the resonator in order to heat it; this causes Rb adsorbed on the surface to be removed. Alternatively, we can heat the resonator by sending 1~mW of power through the coupling fiber. There is also optical access for a long working distance imaging system (Navitar Inc.) used when observing the fiber alignment. The copper block is shielded from external noise sources by viton cushions providing isolation for frequencies above 30~Hz with a suppression of $-20$~dB/decade.

\subsection{Fountain performance}
\begin{figure}[]
\centering
\includegraphics[width=0.45\textwidth]{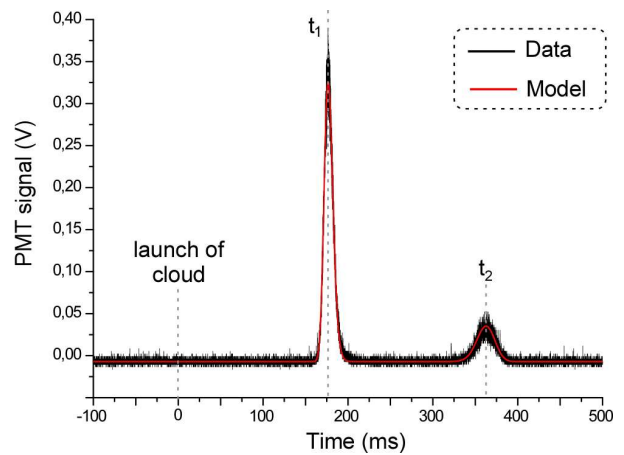}
    \caption{Fluorescence signal of an atomic cloud reaching the probe beam placed at the position of the resonator (with the resonator removed). The data is a concatenation of two measurements: Data around the first peak at $t_1$ shows the cloud reaching the probe beam on its upward trajectory and data around the second peak at $t_2$ shows the cloud falling down after reaching its maximum height. The temperature estimated from a time-of-flight model is around 6~$\mu$K.}
\label{both_peaks_newest}
\end{figure}

The performance of the fountain is analyzed by fluorescence detection of the atomic cloud in the resonator chamber. This allows us to perform time-of-flight measurements in order to determine the atomic velocity and temperature. For this purpose, we launch a light sheet, resonant with the atoms, through the resonator chamber a few centimeters below the resonator position and detect the atomic fluorescence with a photo-multiplier tube (PMT) (Hamamatsu Photonics Deutschland GmbH, H 6780-20). A typical time-of-flight measurement is shown in Fig.~\ref{both_peaks_newest}. Using a simple model that describes the velocity and spatial distribution of the measured signal in terms of temperature, starting velocity, as well as light-sheet beam size and position, several properties can be determined from the data. Based on the width of the peaks it is possible to extract the temperature, which is around 6~$\mu$K. The arrival time of the atomic cloud is checked against the expected arrival time determined from the launch velocity, $v_z$. This enables us to verify the correct operation of the atomic fountain. Details of the model can be found in Ref.~\cite{tesis_Hauswald}, and references therein.

\begin{figure}[]
\centering
\includegraphics[width=0.45\textwidth]{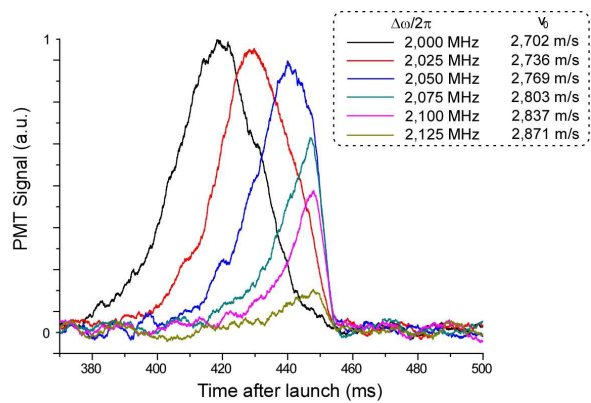}
    \caption{Fluorescence signal of the atomic cloud on its downward trajectory for several different launch velocities. The velocity for each launch was incremented so that a fraction of the cloud hit the top viewport of the resonator chamber and was consequently lost. Only the remaining atoms fall down towards the probe beam. The exact height of the cloud can be controlled by setting the launch velocity, $v_z$, which is determined by the detuning of the cooling beams, $\Delta\omega$, in the MOT chamber.}
\label{cloud_hitting_viewport2}
\end{figure}

Having verified the delivery of cold atoms to the resonator chamber, we next demonstrate the ability to place the cloud at an exact, predetermined height by controlling the launch velocity, $v_z$. The top viewport of the resonator chamber acts as a known reference position. Atoms that collide with the viewport are lost from the cloud and do not contribute to the fluorescence signal which is recorded on the downward trajectory. Figure~\ref{cloud_hitting_viewport2} shows several launches where the turning point of the cloud is incrementally moved closer to the viewport. It can be seen that an increasingly large fraction of the cloud is lost on the viewport, causing the fluorescence signal to be truncated. This demonstrates our ability to control the launch velocity of the cloud and thus to make the turning point of the cloud coincide with resonator position.

\subsection{Atom detection scheme}
The arrival of atoms from the fountain in the evanescent field of the bottle resonator is an inherently random process and experimental sequences must be triggered by these events. For detecting an atom arrival, we plan to tune the resonator into resonance with the D2-transition of $^{85}{\rm Rb}$ and to operate it under the condition of critical coupling while monitoring the transmission of weak detection light through the coupling fiber \cite{Aok06Obs}. This detection light is resonant with the resonator, meaning that its transmission is low when no atom is present. Experimentally, the detection light transmission is at $\sim$1~\% in this case. An atom coupling to the resonator results in new eigenmodes and eigenfrequencies for the atom-resonator system, as described by the Jaynes--Cummings model \cite{Jay63}. For the single photon case, this leads to Rabi splitting of the resonance by an amount equal to the vacuum Rabi frequency, $\Omega_0=2g$. As a result, the transmission of the weak detection field through the coupling fiber will increase when an atom enters the evanescent field of the resonator and will approach unity for the case of strong coupling.

\begin{figure}
\centering\includegraphics[width=0.45\textwidth]{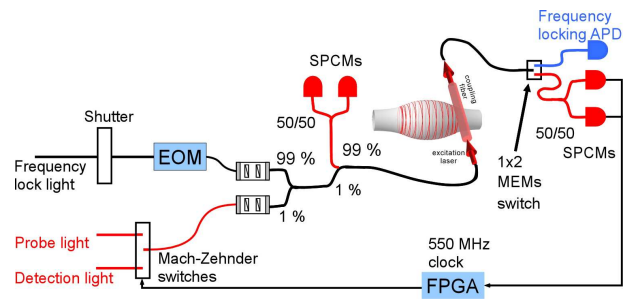}
    \caption{Schematic of the optical experimental setup. The experimental sequence consists of two interlaced phases: first, atoms are cooled and delivered to the resonator, and second, the actual CQED experiment is performed. The frequency of the resonator is actively stabilized during the first phase while the active frequency stabilization is frozen during the second phase. During this second phase, transmission and reflection from the resonator is monitored using single photon counting modules (SPCMs). Realtime detection of atom arrivals is provided by a field programmable field gate array (FPGA) that determines if an atom is strongly coupled to the resonator field. In this case, the detection light is switched off and the probe light is sent towards the resonator.}
\label{CQED_scheme}
\end{figure}

Experimentally, the detection light is optimized for atom detection while a second light field, called probe in the following, is prepared with a preset power and probe--atom detuning, optimized for the actual CQED experiment. Both the detection and the probe light and, in addition, the frequency stabilization light have to be sequentially launched through the coupling fiber. Figure~\ref{CQED_scheme} shows the layout with several fiber couplers for combining and splitting the different optical paths for frequency stabilization, atom detection, and atom probing. Light transmitted past the resonator--fiber coupling junction is either sent to an avalanche photodiode for frequency stabilization of the resonator or, alternatively, to two single photon counting modules (SPCMs) in a Hanbury-Brown and Twiss configuration in order to detect the atom arrival and to record the probe transmission and photon statistics during the actual CQED experiment. Light which is scattered into the anti-clockwise propagating resonator mode is out-coupled in the backward direction and recorded with the aid of two more SPCMs.

The experimental sequence involves operating the frequency stabilization while cooling atoms in the MOT and freezing the stabilization for 85~ms when the atomic cloud reaches the resonator after the fountain launch. During this brief time window, the laser light for stabilization is turned off, the atom detection beam is turned on and the SPCMs are gated on. All transmitted photon detection events are timestamped and recorded on a computer for post-analysis.

The experimental sequence is run several hundred times and is interleaved with an identical reference sequence in which the atoms are not launched. Photon detection events are binned with 1~$\mu$s time resolution for each sequence. The resonator is probed on resonance with an incident power of $P_{\rm in} \approx 2$~pW, corresponding to 0.18 intracavity photons, and the transmission of the empty resonator is maintained at critical coupling with ${P_{\rm out}/P_{\rm in}\leqq0.02}$. Without the cloud being launched, the mean number of photon detection events per time bin is 0.07. Assuming Poissonian statistics, the probability for detecting five photons in one time bin is thus given by $P^{\rm theo}_{\rm ref}(5)=0.9\times10^{-8}$. Evaluating 917 experimental sequences, we find $P^{\rm exp}_{\rm ref}(5)={2.2^{+3.4}_{-1.6}\times10^{-7}}$. The deviation with respect to the theoretical value is due to both frequency noise in the resonator frequency stabilization and mechanical noise in the coupling setup, which both cause super-Poissonian fluctuations of the transmitted intensity. When the atom cloud is launched, however, the photon statistics are significantly modified, yielding a 5-photon detection probability of $P^{\rm exp}_{\rm at}(5)={1.2^{+0.2}_{-0.2}\times10^{-5}}$, almost two orders of magnitude higher than $P^{\rm exp}_{\rm ref}(5)$. The error bars are 95~\% confidence intervals. The observed change in the counting statistics is a clear indication that atoms from the cloud interact with the resonator mode and that this interaction significantly modifies the transmission characteristics of the resonator \cite{Aok06Obs}.

In order to trigger more complex experimental sequences on the arrival of a single atom in the resonator field, the photon detection events have to be evaluated in realtime. For this purpose, the setup incorporates a field programmable gate array (FPGA) with a clock speed of 500~MHz (Xilinx Inc., Virtex-5). The FPGA is programmed to count photon events in a specified time window and discriminates if an atom is strongly coupled to the resonator mode based on the number of counts. When the threshold condition is fulfilled, Mach-Zehnder interferometric switches switch off the detection light and send the probe light towards the resonator.

\section{Summary and outlook}
Summarizing, bottle microresonators provide many advantageous properties for applications in non-linear optics and quantum optics experiments. In particular, in addition to the previous achievements with other types of WGM microresonators, they can be actively stabilized and offer straightforward mechanical access for two ultra-thin coupling fibers. This facilitates their operation as add--drop filters and, in conjunction with their high $Q^2/V$ ratio, allowed us to use them for Kerr-switching of single-wavelength cw light between two fiber ports at record low powers. The ease and efficiency at which these experiments could be performed bodes well for extending such non-linear applications to the quantum regime, where CQED effects with single atoms can be exploited. To this end, we have set up an experimental apparatus compatible with the requirements of cold-atom CQED using bottle microresonators.

\subsection*{Towards CQED experiments with WGM resonators and laser-trapped atoms}
\label{Placing_atom}
WGM microresonators pose a challenge for atom--cavity coupling due to their monolithic geometry. In contrast, the mode of a FP resonator is easily accessible because it is located in free space between mirrors. In Ref.~\cite{Boc04Obs}, a single atom was trapped in a FP microresonator using an intracavity dipole trap created by an auxiliary cavity mode, far detuned from the atomic resonance frequency. A Raman scheme was used in order to further cool the atomic motion and thus to prevent variation in the coupling strength along the cavity axis. For this system, it was possible to measure the Rabi splitting for a single atom placed in the cavity in a single sweep of the probe laser frequency over the resonance of the coupled atom--cavity system.

For WGM resonators, on the other hand, the peak value of the electric field is located inside the resonator material and coupling of an atom to the microresonator mode is only possible via the evanescent field. The spatial decay of the evanescent field amplitude and thus of the coupling strength takes place on a length scale of a fraction of the wavelength. This is much smaller than the accuracy with which a single atom can be positioned by conventional optical trapping techniques. Therefore, up to now, only transits of atoms through the evanescent field were observed for WGM microresonators \cite{Aok06Obs}.

A scheme was proposed in Ref.~\cite{Ver97Qua} where an atom is confined to a circular orbit in close proximity to a microsphere by using two modes, oppositely detuned from the atomic resonance frequency. The red detuned light field creates an attractive potential, whereas the blue detuned light field compensates the vdW attraction close to the surface. Another promising method for placing an atom sufficiently close to the resonator surface is offered by ultra-thin fiber-based dipole traps \cite{Vet10Opt}. In these devices, atoms are trapped in the evanescent field surrounding the fiber waist created by two guided laser fields, again oppositely detuned from the atomic resonance. The strong exponential spatial decay of the evanescent field allows one to control the radial position of the atoms with outstanding accuracy. Critically, this trap is fully compatible with the bottle microresonator concept. The ultra-thin waist of the trapping fiber can be placed at one caustic of the resonator and can, in addition, be used to couple light into and out of the bottle mode used for the CQED experiment.

\begin{acknowledgements}
The authors wish to thank Alexander Rettenmaier, Christian Hauswald, and Sebastian Nickel for their contributions to parts of the presented experimental work. Financial support by the DFG (Research Unit 557), the Volkswagen Foundation (Lichtenberg Professorship), and the ESF (EURYI Award) is gratefully acknowledged.
\end{acknowledgements}


\end{document}